\documentclass[11pt]{article}

\usepackage{graphicx}
\usepackage{amsmath}
\usepackage{dcolumn}
\usepackage{bm}
\usepackage{slashed}
\usepackage[bookmarks,bookmarksnumbered,colorlinks=true,anchorcolor=blue,
linkcolor=blue,urlcolor=blue,citecolor=blue,breaklinks=true]{hyperref}
\usepackage[utf8]{inputenc}
\usepackage{authblk}
\usepackage{cite}
\usepackage{titlesec}
\usepackage{caption}
\usepackage{subcaption}

\usepackage{color}
\usepackage{ulem}

\newcommand{\be}{\begin{equation}}
\newcommand{\ee}{\end{equation}}
\newcommand{\ba}{\begin{eqnarray}}
\newcommand{\ea}{\end{eqnarray}}

\def\bea{\begin{eqnarray}}
\def\eea{\end{eqnarray}}

\newcommand{\gsim}{\mathrel{\hbox{\rlap{\lower.55ex \hbox {$\sim$}}
                   \kern-.3em \raise.4ex \hbox{$>$}}}}
\newcommand{\lsim}{\mathrel{\hbox{\rlap{\lower.55ex \hbox {$\sim$}}
                   \kern-.3em \raise.4ex \hbox{$<$}}}}

\def\roughly#1{\mathrel{\raise.3ex\hbox{$#1$\kern-.75em%
\lower1ex\hbox{$\sim$}}}}
\def\lsim{\roughly<}
\def\gsim{\roughly>}

\def\({\left(}
\def\){\right)}
\def\[{\left[}
\def\]{\right]}
\def\<{\langle}
\def\>{\rangle}

\usepackage{xcolor}

\begin{document}
\title{\bf Entanglement islands read perfect-tensor entanglement}

\author[]{Yi-Yu Lin$^{1,2}$ \thanks{yiyu@bimsa.cn}}
\author[]{Jun Zhang$^{3}$ \thanks{jzhang163@crimson.ua.edu}}
\author[]{Jie-Chen Jin$^{4}$ \thanks{jinjch5@mail2.sysu.edu.cn}}

\affil[]{${}^1$Beijing Institute of Mathematical Sciences and Applications (BIMSA),
	Beijing, 101408, China}
 \affil{${}^2$Yau Mathematical Sciences Center (YMSC), Tsinghua University,
	Beijing, 100084, China}
  \affil{${}^3$Department of Physics and Astronomy, University of Alabama, 514 University Boulevard, Tuscaloosa, AL 35487, USA}
\affil{${}^4$School of Physics and Astronomy, Sun Yat-Sen University, Guangzhou 510275, China} 


\maketitle

\begin{abstract}

In this paper, we make use of holographic Boundary Conformal Field Theory (BCFT)  to simulate the black hole information problem in the semi-classical picture. We investigate the correlation between a portion of Hawking radiation and entanglement islands by the area of an entanglement wedge cross-section. Building on the understanding of the relationship between entanglement wedge cross-sections and perfect tensor entanglement as discussed in reference~\cite{yiyu2023}, we make an intriguing observation: in the semi-classical picture, the positioning of an entanglement island automatically yields a pattern of perfect tensor entanglement. Furthermore, the contribution of this perfect tensor entanglement, combined with the bipartite entanglement contribution, precisely determines the area of the entanglement wedge cross-section.

\end{abstract}

\newpage
\tableofcontents

\section{Background Introduction}
Recently, the holographic AdS/BCFT duality~\cite{Takayanagi:2011zk,Fujita:2011fp,Nozaki:2012qd,Karch:2000gx} has emerged as an intriguing model that elegantly captures the essence of black hole information problems~\cite{Page:1993wv,Page:2013dx}. The story begins with the island formula~\cite{Almheiri:2019hni,Penington:2019npb,Almheiri:2019psf}, which provides a clever interpretation of the peculiar behavior of entanglement entropy in the semi-classical picture when a d-dimensional Conformal Field Theory (CFT) residing on a flat spacetime M is coupled to a gravity theory on a d-dimensional spacetime Q. This formulation offers insights into the Page curve during the black hole evaporation process. More precisely, the correct von Neumann entropy ${S_R}$ (also known as fine-grained entropy) of a subregion R in M is given by the island formula:

\be\label{isl}{S_R} = \mathop {Min\;Ext}\limits_I \left[ {{S_{QFT}}\left( {R \cup I} \right) + \frac{{\text{Area}\left( {\partial I} \right)}}{{4G_N^{\left( d \right)}}}} \right].\ee
Here, I is referred to as the island, which is a region in Q, and $\partial I$ is its spatial boundary. The island formula informs us that, in the semi-classical picture, to compute ${S_R}$, we need to evaluate the field theory entanglement entropy of the region $R \cup I$ (also known as the semi-classical entropy) plus the gravitational area contribution ${S_{{\rm{gravity}}}}$ from the boundary of the island I. The final ${S_R}$ is then determined by minimizing the total contribution. On the other hand, in AdS/BCFT, a d-dimensional holographic BCFT on a manifold M with a boundary $\partial M$ is dual to a d+1-dimensional bulk spacetime N enclosed by an ETW (end of the world) brane Q such that $\partial Q = \partial M$~\cite{Takayanagi:2011zk,Fujita:2011fp}. Crucially, the holographic BCFT also has a third equivalent description purely in terms of a d-dimensional CFT on M coupled to d-dimensional gravity on Q~\cite{Suzuki:2022xwv,Izumi:2022opi}. In the analysis of black hole information problems, this ``triality" is usually qualitatively understood as the combination of AdS/BCFT duality and brane world holography~\cite{Randall:1999ee,Randall:1999vf,Karch:2000ct,Gubser:1999vj}.

~\cite{Suzuki:2022xwv,Izumi:2022opi} explicitly proposes the formulation of the Island/BCFT duality, wherein the gravity on the ETW brane Q is understood as an induced gravity. This induced gravity is described by a d-dimensional CFT coupled to a d-dimensional gravity with its action determined simply by a cosmological constant term. Upon integrating out the CFT fields on Q, one will formally obtain the (minus) Liouville action, which well approximates the effective d-dimensional gravity when the tension of the brane is very large. In the framework of the Island/BCFT duality, the island formula~(\ref{isl}) is elegantly mimicked by the holographic BCFT's Ryu-Takayanagi (RT) formula~\cite{Takayanagi:2011zk,Fujita:2011fp}:
\be\label{bcft}{S_R} = \mathop {Min\;Ext}\limits_{{\Gamma_R}, I} \left[ {\frac{{\text{Area}\left( {{\Gamma_R}} \right)}}{{4G_N^{\left( {d + 1} \right)}}}} \right], \quad \partial {\Gamma_R} = \partial R \cup \partial I.\ee
Here, ${\Gamma_R}$ represents the RT surface corresponding to the subregion R in M (see Fig.~\ref{fig3.1.1}), and I now simulates the island on the ETW brane Q.\footnote{ It is worth mentioning that, due to the fact that now the brane world gravity are purely induced from quantum corrections of matter fields, i.e., all induced gravity contributions are included in the CFT parts, we simply have ${S_{{\rm{gravity}}}} = 0$.}

The island rule is undoubtedly fascinating and surprising. It tells us that, in the semi-classical picture, when calculating the entanglement entropy of a subregion R in a non-gravitational region, we must carefully account for the contribution of the degrees of freedom of a special region I in the gravitational region. A natural interest arises: what is the entanglement structure between different parts of the entire system in such a semi-classical picture? An insightful pictorial understanding is that the island I is actually connected to R through a wormhole in a higher-dimensional spacetime~\cite{Penington:2019kki,Almheiri:2019qdq}. As shown in equation (\ref{bcft}), the island/BCFT duality provides a very concise setup for studying the entanglement structure related to the island, wherein the information of ${S_R}$ is encoded solely in the geometric area of a single bulk extremal surface ${\Gamma_R}$. This provides a convenient quantum information perspective, since it allows us to use the information of a set of extremal surfaces' area to study the entanglement structure related to the island. To be more specific, we will leverage the concept of holographic coarse-grained states proposed recently in~\cite{yiyu2023}, extracting information characterizing the entanglement structure of the holographic d-dimensional system at the coarse-grained level from the area information of extremal surfaces in the d+1-dimensional bulk. This information is characterized by a set of conditional mutual informations (CMIs) and can be visually represented by a set of thread bundles~\cite{Lin:2021hqs,Lin:2022flo,Lin:2022agc,Lin:2023orb}. This kind of pictorial representation using threads originates from the concept of bit threads~\cite{Freedman:2016zud,Cui:2018dyq,Headrick:2017ucz,Headrick:2022nbe}, which were proposed to equivalently formulate the RT formula~\cite{Ryu:2006bv,Ryu:2006ef,Hubeny:2007xt} by convex programming duality. It is worth noting that bit threads, or more generally, the pictorial representation of threads, have insightful significance for the relation between wormholes and quantum entanglement~\cite{Bao:2019wcf,Bao:2021ebo,Rolph:2021hgz}. Furthermore, holographic CMI is closely related to many concepts in the studies of holographic duality~\cite{yiyu2023,Lin:2021hqs,Lin:2022flo,Lin:2022agc,Lin:2023orb,Kudler-Flam:2019oru,Rolph:2021nan}, such as MERA tensor networks~\cite{Vidal:2007hda,Vidal:2008zz,Vidal:2015,Swingle:2009bg,Swingle:2012wq}, kinematic space~\cite{Czech:2015kbp,Czech:2015qta}, holographic entropy cone~\cite{Bao:2015bfa,Hubeny:2018ijt,Hubeny:2018trv,HernandezCuenca:2019wgh}, and holographic partial entanglement entropy~\cite{Vidal:2014aal,Wen:2019iyq,Wen:2018whg,Wen:2020ech,Han:2019scu,Kudler-Flam:2019oru,Lin:2021hqs}.

Since coarse-grained states are conceptually constructed solely from the area information of a set of extremal surfaces in the semi-classical picture, they are expected to provide a characterization of quantum entanglement of the holographic quantum system at the coarse-grained level~\cite{yiyu2023}. The entanglement structure of a genuine holographic quantum system is expected to be much more complex. In fact, we start by simply defining the coarse-grained states constructed directly from a set of CMIs as a class of direct-product states of bipartite entanglement. However, the key point is that, probing entanglement structure from a coarse-grained level can provide insightful clues about the entanglement structure that a holographic system should appear as. ~\cite{yiyu2023} indicates that to further characterize some quantum information theory quantities with geometric duals in holographic duality, such as the entanglement entropy of disconnected regions and entanglement of purification (EoP) dual to the entanglement wedge cross-section (EWCS)~\cite{Nguyen:2017yqw,Takayanagi:2017knl}, perfect tensor entanglement must be introduced, at least at the coarse-grained level, to obtain consistent results.

In this article, following~\cite{yiyu2023}, specifically, based on the understanding of the connection between the entanglement wedge cross-section and perfect tensor entanglement, we attempt to study the entanglement structure when the island appears in the island/BCFT setup. In this setup, we discover a very interesting and noteworthy phenomenon: in the semi-classical picture, for a subregion R in M, the positioning of an entanglement island I, as executed by the island formula (\ref{bcft}), automatically gives rise to a pattern of perfect tensor entanglement between R and I. Moreover, the contribution of perfect tensor entanglement, added to the contribution of bipartite entanglement, precisely gives the area of the entanglement wedge cross-section that characterizes the intrinsic correlation between R and I. In fact, since the entanglement wedge cross-section is the minimal area surface that separates the ``channel’’ connecting R and I, it plays a role in some sense as the horizon of the wormhole (for discussions on this topic, refer to the work~\cite{Bao:2018fso,Bao:2018zab,Bhattacharya:2020ymw}). Although in this paper we restrict ourselves to the context of the simple holographic BCFT models, based on the Island/BCFT duality argued in~\cite{Suzuki:2022xwv,Izumi:2022opi}, we expect that we have captured essential features of the correlation patterns between the island and the ``Hawking radiation’’ R. We anticipate that our study has an inspiring significance for understanding the entanglement patterns in more sophisticated ``radiation-island’’ models (see, e.g.,~\cite{Chen:2020uac,Chen:2020hmv,Akal:2021foz,Almheiri:2019psy,Chen:2019uhq,Balasubramanian:2020hfs,Geng:2020qvw,Bousso:2020kmy,Chen:2020jvn,Chen:2020tes,Akal:2020twv,Miyaji:2021lcq,Geng:2021mic,Bhattacharya:2021nqj,Hu:2022ymx,Anous:2022wqh,Kawamoto:2022etl,Bianchi:2022ulu,Akal:2021dqt,Akal:2022qei,Chang:2023gkt,He:2021mst,Miao:2023unv,Hu:2022zgy,Du:2022vvg,Tong:2023nvi,Li:2023fly,Ling:2020laa,Chu:2021gdb, Deng:2020ent, Basak:2023bnc, Lin:2023ajt, Deng:2023pjs, Basu:2023wmv, Basu:2022crn, Geng:2022dua, Yu:2022xlh, Deng:2022yll, Geng:2021iyq}). This is a natural direction for future work.

The structure of this paper is as follows: In Section\ref{sec2}, we primarily review fundamental concepts used in this work, namely coarse-grained states, the entanglement wedge cross section, and perfect tensor states, as a foundation for the paper. In Section\ref{sec3}, we elaborate on the motivation and proposal of this paper. We highlight a very interesting phenomenon: in the semiclassical picture, the positioning of an entanglement island automatically gives rise to a pattern of perfect tensor entanglement. Furthermore, the sum of this perfect tensor entanglement contribution and the bipartite entanglement contribution precisely gives the area of an EWCS. In Section\ref{sec4}, we validate our proposal in various scenarios, including finite-size subregions in a semi-infinite straight line BCFT, subregions on a disk BCFT, and BCFT setups simulating the black hole information problem. The final section is the conclusion and discussion.


\section{Review: Coarse-Grained States, Entanglement Wedge Cross-Section, and Perfect Tensor State}\label{sec2}

This section mainly reviews the basic concepts needed in this article: coarse-grained states, entanglement wedge cross-section, and perfect tensor states, laying the foundation for the discussion.
\subsection{Thread Configurations and Coarse-Grained States}
\begin{figure}
     \centering
     \begin{subfigure}[b]{0.4\textwidth}
         \centering
         \includegraphics[width=\textwidth]{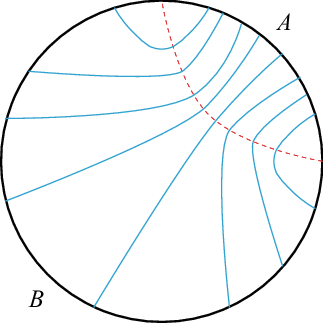}
         \caption{}
         \label{2.1.1a}
     \end{subfigure}
     \hfill
     \begin{subfigure}[b]{0.4\textwidth}
         \centering
         \includegraphics[width=\textwidth]{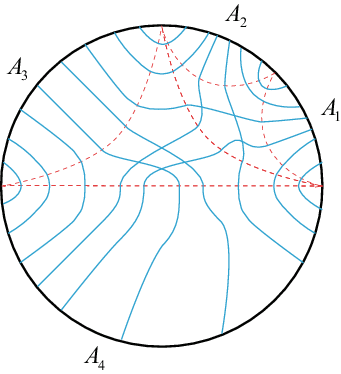}
         \caption{}
         \label{2.1.1b}
     \end{subfigure}
     \caption{(a) A ``thread'' picture characterizing the entanglement entropy between two complementary regions. (b) A more refined thread configuration characterizing a set of entanglement entropies involving more subregions. Here the threads are schematically represented as blue lines, and the RT surfaces are represented as red dashed lines.}
\end{figure}

We first review some important conclusions from~\cite{yiyu2023}. In~\cite{yiyu2023}, a coarse-grained state is proposed as a quantum state characterizing the entanglement structure of a holographic quantum system at the coarse-grained level, constructed from a sets of CMIs, and simply a direct product of Bell states characterizing bipartite entanglement. These states can be pictorially represented using a series of thread configurations. In the framework of AdS/CFT duality~\cite{Maldacena:1997re,Gubser:1998bc,Witten:1998qj}, we can naively imagine a pictorial representation of the entanglement structure revealed by the RT formula~\cite{Ryu:2006bv,Ryu:2006ef,Hubeny:2007xt}, as shown in the figure~\ref{2.1.1a}. Considering the entanglement entropy between a subsystem A and its complement B in a pure state of the holographic CFT, one can imagine a family of uninterrupted threads, with each end connected to A and its complement B, respectively, passing through the RT surface ${\gamma_A}$. The number of threads ${F_{AB}}$ precisely equals the entanglement entropy $S(A)$ between A and B:
\be{F_{AB}} = {S_A}.\ee
These threads are commonly referred to as ``bit threads’’~\cite{Freedman:2016zud,Cui:2018dyq,Headrick:2017ucz,Headrick:2022nbe}. It is natural to further construct a series of finer thread configurations~\cite{Headrick:2020gyq,Lin:2020yzf,Lin:2021hqs,Lin:2022flo,Lin:2022agc}, allowing us to calculate the entanglement entropy for more than one region. As shown in Fig.~\ref{2.1.1b}, we can further decompose region A into $A = {A_1} \cup {A_2}$ and B into $B = {A_3} \cup {A_4}$. Then, we can construct a finer thread configuration that calculates the entanglement entropy between six connected regions and their complements, including: $A_1$, $A_2$, $A_3$, $A_4$, as well as $A = {A_1} \cup {A_2}$, ${A_2} \cup {A_3}$. In other words, the number of threads connecting one of these six regions and its complement precisely equals its corresponding entanglement entropy. Similarly, one can further divide the quantum system M into more and more adjacent and non-overlapping basic regions ${A_1}, {A_2}, \ldots, {A_N}$, and then obtain correspondingly finer thread configurations. Note that here we have carefully drawn the threads to appear perpendicular to the RT surfaces they pass through, in keeping with the conventional property of bit threads. However, in the sense of coarse-grained states, only the topology is really important, and we have not yet seriously considered the exact trajectories of these threads. Nevertheless, this process can be iterated as long as each basic region remains much larger than the Planck length to ensure the applicability of the RT formula. Here we define basic regions satisfying~\cite{Lin:2021hqs,Lin:2022flo,Lin:2022agc}:
\be{A_i} \cap {A_j} = \emptyset, \;\, \cup {A_i} = M.\ee
Then for each pair of basic regions ${A_i}$ and ${A_j}$, we define a function ${F_{ij}} \equiv {F_{{A_i} \leftrightarrow {A_j}}}$, representing the number of threads connecting ${A_i}$ and ${A_j}$. The generalized finer thread configuration corresponding to $\left\{ {{F_{ij}}} \right\}$ satisfies:
\be\label{equ}{S_{a(a + 1) \ldots b}} = \sum\limits_{i,j} {{F_{ij}}}, \quad {\rm{where}}\;i \in \left\{ {a, a + 1, \ldots, b} \right\}, \;j \notin \left\{ {a, a + 1, \ldots, b} \right\}.\ee
Here, ${S_{a(a + 1) \ldots b}}$ represents the entanglement entropy of a connected composite region $A = {A_{a(a + 1) \ldots b}} \equiv {A_a} \cup {A_{a + 1}} \cup \ldots \cup {A_b}$. This equation can be intuitively understood as the entanglement entropy between A and $\bar{A}$ arising from the sum of ${F_{ij}}$ for basic regions ${A_i}$ inside A and basic regions ${A_j}$ inside the complement $\bar{A}$. Thread configurations that satisfy condition (\ref{equ}) are commonly referred to as ``locking’’ thread configurations, borrowing the term from network flow theory. Solving condition~(\ref{equ}), it turns out that the number of threads connecting two basic regions is precisely given by the so-called conditional mutual information. In other words, the conditional mutual information characterizes the correlation between two regions $A_i$ and $A_j$ separated by a distance L:
\be\label{cmi}\begin{array}{l}
{F_{{A_i} \leftrightarrow {A_j}}} = \frac{1}{2}I({A_i}, {A_j}|L)\\
\quad \quad \;\;\,{\kern 1pt}  \equiv \frac{1}{2}[S({A_i} \cup L) + S({A_j} \cup L) - S({A_i} \cup L \cup {A_j}) - S(L)]
\end{array}.\ee
Here, we denote the region in the middle of ${A_i}$ and ${A_j}$ as $L = {A_{\left(i + 1\right) \cdots \left(j - 1\right)}}$, which is a composite region composed of many basic regions and also represents the distance between ${A_i}$ and ${A_j}$.

\begin{figure}
    \centering
    \includegraphics{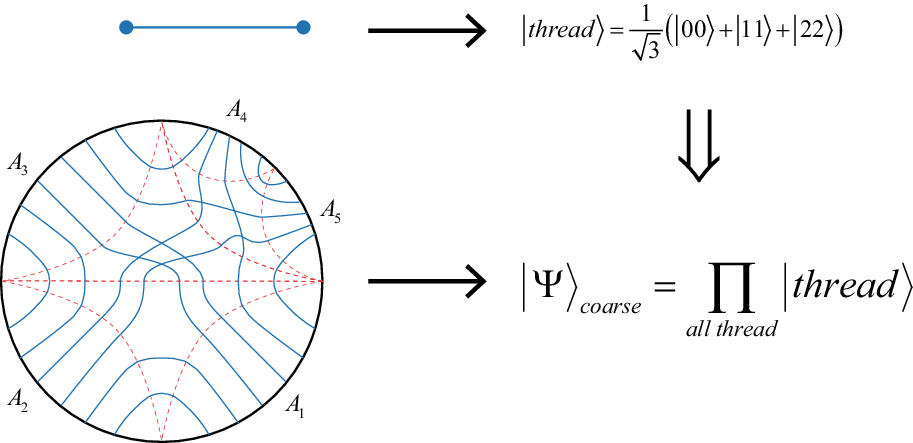}
    \caption{The refined thread configurations characterizing a set of CMIs should be understood as a kind of coarse-grained states, which only characterize the quantum entanglement of the holographic quantum system at a coarse-grained level. }
    \label{state}
\end{figure}

In fact, this kind of refined thread configuration is closely related to various concepts proposed from different perspectives in holographic duality research, including kinematic space~\cite{Czech:2015kbp,Czech:2015qta}, entropy cone~\cite{Bao:2015bfa,Hubeny:2018ijt,Hubeny:2018trv,HernandezCuenca:2019wgh}, holographic partial entanglement entropy~\cite{Vidal:2014aal,Wen:2019iyq,Wen:2018whg,Wen:2020ech,Han:2019scu,Kudler-Flam:2019oru,Lin:2021hqs}, and MERA tensor networks~\cite{Vidal:2007hda,Vidal:2008zz,Vidal:2015,Swingle:2009bg,Swingle:2012wq}. These connections are, in some sense, natural and easily obtained, especially where conditional mutual information plays a central role, defined as the volume measure in kinematic space, characterizing the density of entanglement entropy. Discussions on these connections can be found in a series of articles~\cite{yiyu2023,Lin:2021hqs,Lin:2022flo,Lin:2022agc,Lin:2023orb,Kudler-Flam:2019oru,Rolph:2021nan}. The key point is that, within our framework, we will understand these refined thread configurations as a kind of coarse-grained state of the holographic quantum system. These coarse-grained states only characterize the quantum entanglement of the holographic quantum system at a coarse-grained level. In simple terms, this idea suggests that, in these locking thread configurations, each thread can be understood as a pair of maximally entangled qudits. One end of the thread corresponds to a qudit. For example, let us take $d=3$, so one end of the thread corresponds to a qutrit. Thus, each thread corresponds to~\footnote{In~\cite{Lin:2022flo,Lin:2022agc,Lin:2023orb}, this is usually referred to as the ``thread-state’’ duality. In fact, the idea of the thread-state duality conveys more meaning than what is presented by (\ref{qut}). Each thread actually represents not only the entanglement between the two endpoints of the thread but also the entanglement between all degrees of freedom traversed by the thread within the holographic bulk. More details can be found in~\cite{Lin:2022flo,Lin:2022agc,Lin:2023orb}.}
\be\label{qut}\left| {\text{thread}} \right\rangle  = \frac{1}{{\sqrt{3}}}\left( {\left| {00} \right\rangle  + \left| {11} \right\rangle  + \left| {22} \right\rangle } \right).\ee
Then the direct product of states (\ref{qut}) of all threads in the thread configuration gives a coarse-grained state of the quantum system:
\be\label{coar}{\left| \Psi \right\rangle _{\text{coarse}}} = \prod\limits_{\text{all threads}} {\left| {\text{thread}} \right\rangle }.\ee
It can be verified that if we take the partial trace of this coarse-grained state ${\left| \Psi \right\rangle _{\text{coarse}}}$ to obtain the reduced density matrices of each connected region $\left\{ {A_i, A_i A_{i + 1}, A_i A_{i + 1} A_{i + 2}, \ldots} \right\}$, and calculate the corresponding von Neumann entropy, we exactly obtain a set of correct holographic entanglement entropies~\cite{Lin:2022flo,Lin:2022agc,Lin:2023orb}. As the name suggests, coarse-grained states only characterize the quantum entanglement of the holographic quantum system at a coarse-grained level. The entanglement structure of a genuine holographic quantum system is expected to be much more complex. The point is that studying these coarse-grained states will lead us to discover some properties of the entanglement structure in holographic quantum systems.

\subsection{Entanglement Wedge Cross-Section and Perfect Tensor State}\label{sec22}

\begin{figure}
     \centering
     \begin{subfigure}[b]{0.4\textwidth}
         \centering
         \includegraphics[width=\textwidth]{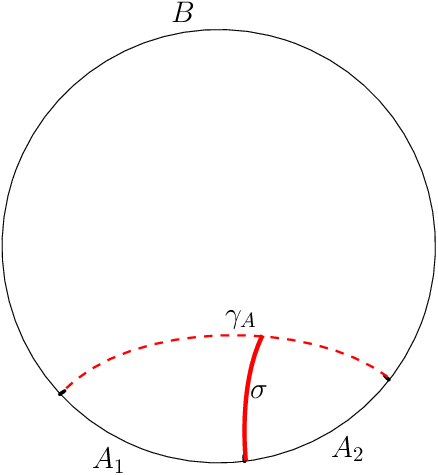}
         \caption{}
         \label{fig2.2.1a}
     \end{subfigure}
     \hfill
     \begin{subfigure}[b]{0.4\textwidth}
         \centering
         \includegraphics[width=\textwidth]{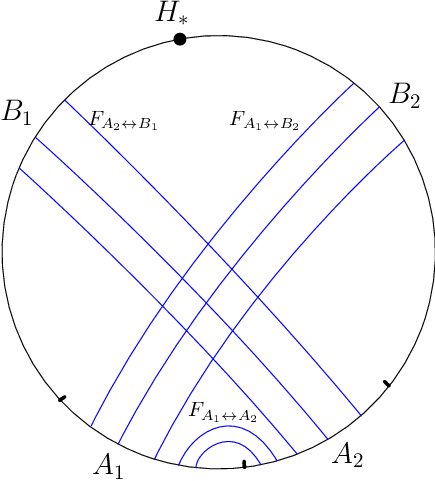}
         \caption{}
         \label{fig2.2.1b}
     \end{subfigure}
     \caption{(a)	The correlation between two adjacent subregions $A_1$ and $A_2$ can be holographicly measured by the area of the entanglement wedge cross section ${\sigma _{{A_1}:{A_2}}}$, represented by the purple line. (b) The BPE method to calculate the area of ${\sigma _{{A_1}:{A_2}}}$: divide the complement $B$ into two parts such that ${F_{{A_1} \leftrightarrow {B_2}}} = {F_{{A_2} \leftrightarrow {B_1}}}$.}
\end{figure}

~\cite{yiyu2023} points out a noteworthy phenomenon regarding the holographic entanglement wedge cross-section, indicating the inevitability of perfect tensor entanglement for the coarse-grained entanglement structure of holographic quantum systems.

Within the framework of AdS/CFT duality, the correlation between two adjacent subregions $A_1$ and $A_2$ in the CFT can be measured by the area of a so-called entanglement wedge cross-section $\sigma_{{A_1}:{A_2}}$~\cite{Nguyen:2017yqw,Takayanagi:2017knl}. The definition of the entanglement wedge cross-section $\sigma_{{A_1}:{A_2}}$ is as follows: first, define the entanglement wedge $W(A)$ of $A = {A_1} \cup {A_2}$ as the bulk region enclosed by A and its corresponding RT surface $\gamma_A$. Then, a minimal area extremal surface $\sigma_{{A_1}:{A_2}}$ is defined, satisfying: 1. Dividing the entanglement wedge $W(A)$ of A into two parts, one entirely touching ${A_1}$ and the other entirely touching ${A_2}$. 2. Selecting the extremal surface with the smallest area among all surfaces satisfying condition 1. 

~\cite{Wen:2021qgx} introduces an interesting method, referred to as the Balanced Partial Entanglement (BPE) method, to holographically calculate the area of this surface, although it is formulated in the language of partial entanglement entropy~\cite{Vidal:2014aal}. We paraphrase it in our thread language as follows. First, we look for a point H on the complement B of A and divide B into $B_1$ and $B_2$, accordingly one can construct a locking thread configuration corresponding to the choice of basic regions $\left\{ {{A_1},\;{A_2},\;{B_1},\;{B_2}} \right\}$. The requirement is to find a point $H_*$ such that, in its corresponding thread configuration, the number of threads connecting $A_1$ and $B_2$ is equal to the number of threads connecting $A_2$ and $B_1$:
\be {F_{{A_1} \leftrightarrow {B_2}}} = {F_{{A_2} \leftrightarrow {B_1}}} ,\ee
Then, it turns out that
\be\label{sig} \frac{{\text{Area}\left( {\sigma_{{A_1}:{A_2}}} \right)}}{{4G_N}} = {F_{{A_1} \leftrightarrow {B_2}}} + {F_{{A_1} \leftrightarrow {A_2}}}. \ee
Explaining the fact that the sum of CMIs can be used to probe this geometric area is crucial. Essentially, the area of the $\sigma_{{A_1}:{A_2}}$ surface measures the correlation between the two parts ${A_1}$ and ${A_2}$ within A. In previous research, this correlation has been proposed to be understood as several quantum information theory quantities such as entanglement of purification (EoP)~\cite{Nguyen:2017yqw,Takayanagi:2017knl}, reflected entropy~\cite{Dutta:2019gen}, logarithmic negativity~\cite{Kudler-Flam:2018qjo,Kusuki:2019zsp}, odd entropy~\cite{Tamaoka:2018ned}, balanced partial entanglement (BPE)~\cite{Wen:2021qgx,Camargo:2022mme,Wen:2022jxr}, differential purification~\cite{Espindola:2018ozt}, and so on. ~\cite{yiyu2023} suggests that when applying coarse-grained states to understand this ``experimental fact’’, the role of perfect tensor entanglement must inevitably be introduced into the coarse-grained entanglement structure of the holographic system. The idea is that one must ``tie’’ the threads connecting $A_1$ and $B_2$ with the threads connecting $A_2$ to $B_2$ to form a special entangled state for four qutrits, identified as a so-called perfect tensor state~\cite{Pastawski:2015qua,Helwig:2012nha,Helwig:2013qoq}: 
\be\label{a1b2}\begin{array}{l}
\left| {{a_1}{b_2}{a_2}{b_1}} \right\rangle  = \frac{1}{3}(\left| {{0_{{a_1}}}{0_{{b_2}}}{0_{{a_2}}}{0_{{b_1}}}} \right\rangle  + \left| {{1_{{a_1}}}{1_{{b_2}}}{1_{{a_2}}}{0_{{b_1}}}} \right\rangle  + \left| {{2_{{a_1}}}{2_{{b_2}}}{2_{{a_2}}}{0_{{b_1}}}} \right\rangle \\
\quad \quad \quad \quad \;\,\,\, + \left| {{0_{{a_1}}}{1_{{b_2}}}{2_{{a_2}}}{1_{{b_1}}}} \right\rangle  + \left| {{1_{{a_1}}}{2_{{b_2}}}{0_{{a_2}}}{1_{{b_1}}}} \right\rangle  + \left| {{2_{{a_1}}}{0_{{b_2}}}{1_{{a_2}}}{1_{{b_1}}}} \right\rangle \\
\quad \quad \quad \quad \,\;\,\, + \left| {{0_{{a_1}}}{2_{{b_2}}}{1_{{a_2}}}{2_{{b_1}}}} \right\rangle  + \left| {{1_{{a_1}}}{0_{{b_2}}}{2_{{a_2}}}{2_{{b_1}}}} \right\rangle  + \left| {{2_{{a_1}}}{1_{{b_2}}}{0_{{a_2}}}{2_{{b_1}}}} \right\rangle )
\end{array}.\ee

As shown in the figure~, once we handle the coarse-grained state in this way, we can clearly see that the correlation between $A_1$ and $A_2$ is precisely composed of two parts. One part (the second term in (\ref{sig})) is contributed by the bipartite entanglement
\be\label{a1a2}\left| {{a_1}{a_2}} \right\rangle  = \frac{1}{{\sqrt 3 }}(\left| {{0_{{a_1}}}{0_{{a_2}}}} \right\rangle  + \left| {{1_{{a_1}}}{1_{{a_2}}}} \right\rangle  + \left| {{2_{{a_1}}}{2_{{a_2}}}} \right\rangle ),\ee
with log3 of entanglement between $a_1$ and $a_2$. The other part (the first term in (\ref{sig})) is contributed by the entanglement of the perfect state (\ref{a1b2}), in which $a_1$ is entangled with the other three qutrits in its complement, resulting in log3 of correlation between $a_1$ and $a_2$. It is not difficult to understand the necessity of the perfect state entanglement: imagine if we only use bipartite entanglement, i.e., if we do not ``tie’’ the threads connecting $A_1$ and $B_2$ with the threads connecting $A_2$ to $B_2$ to create entanglement, then there will be no correlation between $a_1$ and $a_2$ at this point, as seen in its corresponding state
\be\label{pro}\left| {{a_1}{b_2}{a_2}{b_1}} \right\rangle  = \frac{1}{{\sqrt 3 }}(\left| {{0_{{a_1}}}{0_{{b_2}}}} \right\rangle  + \left| {{1_{{a_1}}}{1_{{b_2}}}} \right\rangle  + \left| {{2_{{a_1}}}{2_{{b_2}}}} \right\rangle ) \otimes \frac{1}{{\sqrt 3 }}(\left| {{0_{{a_2}}}{0_{{b_1}}}} \right\rangle  + \left| {{1_{{a_2}}}{1_{{b_1}}}} \right\rangle  + \left| {{2_{{a_2}}}{2_{{b_1}}}} \right\rangle ).\ee
Because at this point, the two threads $a_1b_2$ and $a_2b_1$ are independent. In this way, we would miss the first contribution in (\ref{sig}) and fall into a contradiction. The key point is that the naive state (\ref{pro}) is not completely symmetric about the four qutrits. Note that in it, the entanglement entropy between qutrits $a_1$ union $a_2$ with $b_1$ union $b_2$ is $2\log 3$, while the entanglement entropy between qutrits $a_1$ union $b_2$ and $a_2$ union $b_1$ is 0. On the other hand, the perfect tensor state (\ref{a1b2}) is completely symmetric about the four qutrits. One can verify that this state has an interesting feature: for any qutrit of the four, its entanglement entropy with its complement is log3, and for any two qutrits of the four, the entanglement entropy is $2\log 3$. Perfect tensor states are famous in the holographic HaPPY code model~\cite{Pastawski:2015qua}. They are used as important ingredients for quantum error correction in quantum information theory, also known as Absolutely Maximally Entangled (AME) states~\cite{Helwig:2012nha,Helwig:2013qoq}. More generally, a perfect tensor can be defined in two equivalent ways:

\textbf{Definition 1:} A $2s$-perfect tensor is a $2s$-qudit pure state for any positive integer $s$ such that the reduced density matrix involving any $s$ qudits is maximally mixed.

\textbf{Definition 2:} A $2s$-perfect tensor is a $2s$-qudit pure state for any positive integer $k \leq s$ such that the mapping from the state of any $k$ qudits to the state of the remaining $2s - k$ qudits is an isometric isomorphism.

Furthermore, by replacing direct-product states with perfect tensor states, it is also possible to use coarse-grained states to characterize the entanglement entropy for disconnected regions. For more detailed discussion, refer to~\cite{yiyu2023}.



\section{ Motivation and Proposal}\label{sec3}

\subsection{Semi-Classical Picture of the BCFT on an Infinite Straight Line}
\begin{figure}
    \centering
    \includegraphics{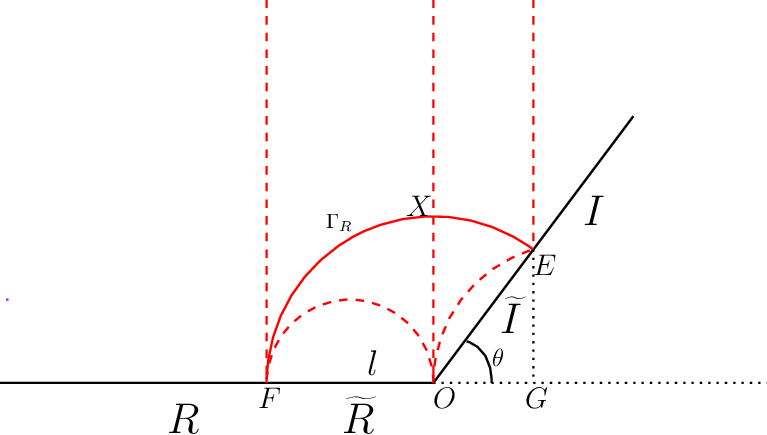}
    \caption{A holographic BCFT on half-space $M$ $x < 0$ and the corresponding ETW brane $Q$. ${\Gamma_R}$ (the red solid line) represents the RT surface corresponding to the subregion R in M, and $I$ is the island region on $Q$. A set of minimal bulk surfaces correspond to the semi-classical entropies are represented by red dashed lines.}
    \label{fig3.1.1}
\end{figure}

To succinctly elucidate our motivation and proposal, let us start with a holographic BCFT in d=2, keeping it as simple as possible. Using the Poincaré coordinate:
\be
ds^2 = L^2 \frac{-dt^2 + dz^2 + dx^2}{z^2}.
\ee
We define a BCFT on half-space M represented by $x < 0$. Figure~\ref{fig3.1.1} illustrates a time slice of this setup. To compute the minimal surface using (\ref{bcft}), the first step is to determine the position of the Q-brane. According to~\cite{Takayanagi:2011zk,Fujita:2011fp}, the gravity action for the AdS/BCFT duality is given by:
\be
I_G = \frac{1}{16\pi G_N} \int_N \sqrt{-g} (R - 2\Lambda) + \frac{1}{8\pi G_N} \int_Q \sqrt{-h} (K - T).
\ee
Here, $h^{ab}$ is the induced metric on the Q-brane, and the constant T is the tension of the Q-brane. Variation of this action leads to a Neumann boundary condition for the Q-brane, determining its position as a plane:
\be
z = \lambda x,
\ee
where
\be\label{lam}
\lambda = \sqrt{\left(\frac{d-1}{LT}\right)^2 - 1}.
\ee
As shown in the Fig.\ref{fig3.1.1}, the simplest case is considering a point F in the BCFT system, dividing the half-line BCFT into two regions, denoted as R and $\tilde{R}$, where R extends to infinity, and $\tilde{R}$ is adjacent to the boundary of BCFT. Let's carefully denote the boundary degrees of freedom of BCFT as B, not included in $\tilde{R}$, but holographically dual to the degrees of freedom of Q. From the perspective of BCFT, our interest lies in the entanglement entropy between region R and $\tilde{R} \cup B$, denoted as $S_R$. However, according to the island/BCFT duality~\cite{Suzuki:2022xwv,Izumi:2022opi} and the island formula (\ref{bcft}), it should be understood as the field theory entanglement entropy of the union of R and the island I in the semi-classical picture:
\be
S_R = S_{\text{semi}}(R \cup I),
\ee
where $I$ is shown in the figure~\ref{fig3.1.1}, determined by the RT surface corresponding to $R$ in the AdS/BCFT duality. Specifically, the RT formula guides us to find a minimal surface in the bulk, where the endpoints, in the context of holographic BCFT, can lie on the ETW brane. This surface should partition the bulk into two parts, one entirely touching the selected boundary subregion R, and the other entirely touching the complement of R. Let us denote the two parts on the Q-brane as I and $\tilde{I}$, adhering to R and $\tilde{R}$ respectively. In this simple context, the RT surface $\Gamma_R$ corresponding to R is well-known, identified as the minimal geodesic EF connecting a point E on the Q-brane and F. The coordinates of point E are given as~\cite{Takayanagi:2011zk,Fujita:2011fp} 
\be\label{eco}
E = (t = 0, x = l\cos \theta, z =l \sin \theta).
\ee

\begin{figure}
     \centering
     \begin{subfigure}[b]{0.5\textwidth}
         \centering
         \includegraphics[width=\textwidth]{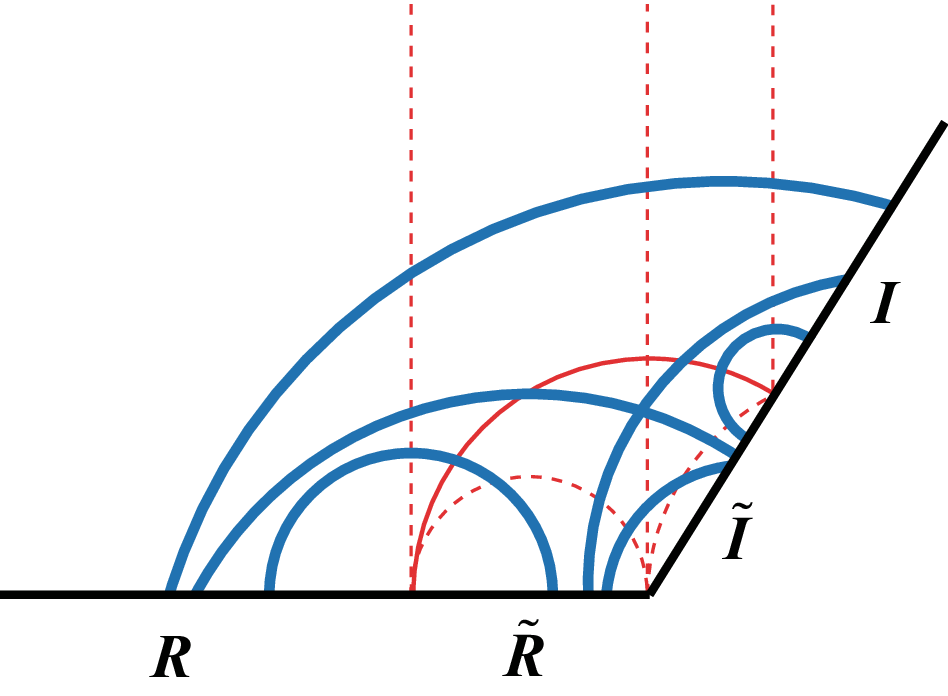}
         \caption{}
         \label{thread}
     \end{subfigure}
     \hfill
     \begin{subfigure}[b]{0.35\textwidth}
         \centering
         \includegraphics[width=\textwidth]{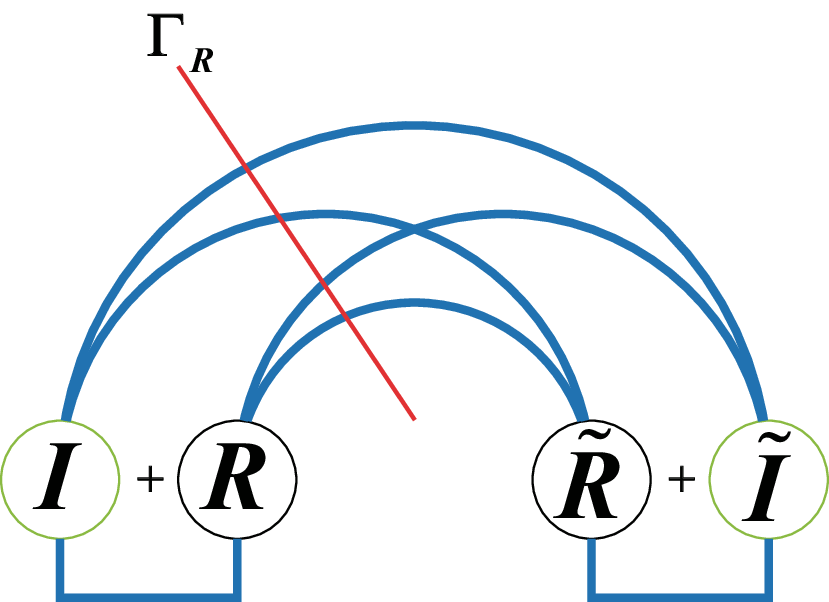}
         \caption{}
         \label{group}
     \end{subfigure}
     \caption{(a) A locking thread configuration corresponding to the coarse-grained level $R \cup \tilde{R} \cup \tilde{I} \cup I$, consisting of six independent thread bundles. Each bundle is represented by a thick blue line. (b) When computing $S_R$, it is as if we have divided the entire system into two groups, $\{\tilde{R}, \tilde{I}\}$ and $\{R, I\}$, and $S_R$ is actually calculating the entanglement entropy between these two groups.}
\end{figure}

From now on, following the idea of island/BCFT, we focus on the semi-classical perspective, where this system is described as a CFT on M coupled to an induced gravity on Q. Let's denote the interface as O. Thus, the points F, O, E partition the system $M \cup Q$ into four parts R, $\tilde{R}$, I, and $\tilde{I}$. We are going to consider these four regions as basic regions and study the entanglement structure of the system on this coarse-grained level. The necessary mathematical results have been derived in our previous work~\cite{Lin:2022aqf}, although the motivation and considerations were based on bit threads and partial entanglement entropy. As shown in the figure~\ref{thread}, we first assign a corresponding locking thread configuration to the coarse-grained system $R \cup \tilde{R} \cup \tilde{I} \cup I$, consisting of six independent thread bundles. For simplicity, each bundle is represented by a thick blue line in the figure, and the number of threads in each bundle is denoted as ${F_{R\tilde{R}}}$, ${F_{RI}}$, ${F_{R\tilde{I}}}$, ${F_{\tilde{RI}}}$, ${F_{\tilde{R}\tilde{I}}}$, and ${F_{I\tilde{I}}}$, which is exactly half of the conditional mutual information, as per (\ref{cmi}). Let's transcribe it as follows:
\be\label{fii}
\begin{bmatrix}
    {F_{I\tilde{I}}} \\
    {F_{\tilde{R}\tilde{I}}} \\
    {F_{R\tilde{I}}} \\
    {F_{\tilde{RI}}} \\
    {F_{R\tilde{R}}} \\
    {F_{RI}}
\end{bmatrix} =
\begin{bmatrix}
    \frac{1}{2}\left(S(\tilde{I}) + S(I) - S(R\tilde{R})\right) \\
    \frac{1}{2}\left(S(\tilde{I}) + S(\tilde{R}) - S(\tilde{I}\tilde{R})\right) \\
    \frac{1}{2}\left(S(\tilde{I}\tilde{R}) + S(R\tilde{R}) - S(\tilde{R}) - S(I)\right) \\
    \frac{1}{2}\left(S(\tilde{I}\tilde{R}) + S(R\tilde{R}) - S(\tilde{I}) - S(R)\right) \\
    \frac{1}{2}\left(S(\tilde{R}) + S(R) - S(R\tilde{R})\right) \\
    \frac{1}{2}\left(S(R) + S(I) - S(\tilde{I}\tilde{R})\right)
\end{bmatrix}
\ee
Here are some comments on this equation set. Firstly, a shorthand notation is used, omitting the union symbols, for example, $R \cup \tilde{R} = R\tilde{R}$. Secondly, the entropy involved here refers to the semi-classical entropy, i.e., the field theory entropy in a fixed background. The usual subscripts ``semi’’ or ``QFT’’ are omitted for brevity, and this notation will be used  throughout this paper. Thirdly, note that on the right side of (\ref{fii}), we only involve the values of six independent entropies, corresponding to the six connected subregions $\{R, \tilde{R}, I, \tilde{I}, \tilde{R}\tilde{I}, R\tilde{R}\}$. According to the spirit of the RT formula, these entropies are precisely calculated by six corresponding RT surfaces, denoted as \{$\gamma(R), \gamma(\tilde{R}), \gamma(I), \gamma(\tilde{I}), \gamma(\tilde{R}\tilde{I}), \gamma(R\tilde{R})\}$, which are depicted by red dashed lines in the figure~\ref{thread}. Fourthly, note that compared to (\ref{cmi}), we have implicitly used an assumption in (\ref{fii}) that M and Q are in a pure state as a whole. Therefore, the entanglement entropy of a region is equal to the entanglement entropy of its complement. For example, $S(\tilde{I}\tilde{R}) = S(IR) \equiv S_R$.

It can be verified that the thread configuration characterizing the CMI fluxes (\ref{fii}) is sufficient to describe the entanglement structure on a coarse-grained level of the system $M \cup Q$ in the semi-classical picture. By plugging in the equation (\ref{equ}), these fluxes automatically provide the entanglement entropies of the six connected subregions \{$R, \tilde{R}, I, \tilde{I}, \tilde{R}\tilde{I}, R\tilde{R}$\}:
\be\label{rec}
\begin{array}{l}
{F_{R\tilde{R}}} + {F_{R\tilde{I}}} + {F_{\tilde{RI}}} + {F_{I\tilde{I}}} = S(\tilde{R}\tilde{I}) = S_R \\
{F_{R\tilde{R}}} + {F_{RI}} + {F_{R\tilde{I}}} = S(R) \\
{F_{R\tilde{R}}} + {F_{\tilde{RI}}} + {F_{\tilde{R}\tilde{I}}} = S(\tilde{R}) \\
{F_{RI}} + {F_{\tilde{RI}}} + {F_{I\tilde{I}}} = S(I) \\
{F_{R\tilde{I}}} + {F_{\tilde{R}\tilde{I}}} + {F_{I\tilde{I}}} = S(\tilde{I}) \\
{F_{RI}} + {F_{R\tilde{I}}} + {F_{\tilde{RI}}} + {F_{\tilde{R}\tilde{I}}} = S(R\tilde{R})
\end{array}.
\ee
The first equation in (\ref{rec}) indicates an insightful feature brought by the thread picture~\cite{Lin:2022aqf}: the contribution to the fine-grained entropy or von Neumann entropy $S_R$ actually comes from four types of entanglement, namely ${F_{R\tilde{R}}}$, ${F_{R\tilde{I}}}$, ${F_{\tilde{RI}}}$, and ${F_{I\tilde{I}}}$, as shown in the figure~\ref{group}. This phenomenon leads to an understanding: when computing $S_R$, it is as if we have divided the entire system into two groups, $\{\tilde{R}, \tilde{I}\}$ and $\{R, I\}$, and $S_R$ is actually calculating the entanglement entropy between these two groups. Therefore, ${F_{\tilde{R}\tilde{I}}}$ and ${F_{RI}}$ cannot contribute to the entanglement entropy between these two groups because they represent the internal entanglement. This aligns well with the meaning of the island rule in the island/BCFT context, i.e., when we want to compute the true von Neumann entropy of R in the semiclassical picture, we also have to consider a somewhat unexpected region I as part of the same group.

Using the geometric dual of entanglement entropy, we can determine the specific values of the fluxes for each thread bundle, by calculating the lengths of the geodesics associated with the six connected regions and plugging them into the formula (\ref{fii}), as detailed in~\cite{Lin:2022aqf}.

\subsection{A Coincidence of two Equal Fluxes}

We are interested in a new question that deepens our understanding of the entanglement structure of the island phenomenon at a coarse-grained level. The motivation arises from a coincidence. When we specifically calculate the numerical values in (\ref{fii}), in particular, we find that
\be\label{iti}{F_{R\tilde I}} = {\textstyle{1 \over 2}}\left( {S\left( {\tilde I\tilde R} \right) + S\left( {R\tilde R} \right) - S\left( {\tilde R} \right) - S\left( I \right)} \right) = \frac{L}{{8G_N^{\left( {d + 1} \right)}}}\ln (2(1 + \cos \theta )),\ee
\be\label{rti}{F_{\tilde RI}} = {\textstyle{1 \over 2}}\left( {S\left( {\tilde I\tilde R} \right) + S\left( {R\tilde R} \right) - S\left( {\tilde I} \right) - S\left( R \right)} \right) = \frac{L}{{8G_N^{\left( {d + 1} \right)}}}\ln (2(1 + \cos \theta )).\ee
Thus, we coincidentally find that
\be\label{coi}{F_{R\tilde I}} = {F_{\tilde RI}},\ee
where we use several semiclassical entropies calculated by the RT formula as follows. The first entropy is calculated by (\ref{bcft}):
\be\label{sir}{S_R} = S(\tilde I\tilde R) = \frac{1}{{4G_N^{\left( {d + 1} \right)}}}{d_{EF}} = \frac{L}{{4G_N^{\left( {d + 1} \right)}}}(\ln \frac{{2l}}{\varepsilon } + \ln \frac{{(1 + \cos \theta )}}{{\sin \theta }}),\ee
where ${d_{EF}}$ represents the length of the geodesic connecting F and the optimized point E. l is the size of the region $\tilde R$, $\varepsilon $ is the UV cutoff, and $\theta$ is the angle between the membrane Q and M, related to the tension T of the brane by (\ref{lam}) through $\tan \theta  = \lambda $ \footnote{In practice, we implicitly consider the setups with $\theta  \approx 0$, corresponding to a very large tension of the Q brane.}. Note that the second term in (\ref{sir}) is holographically dual to the boundary entropy ${S_{bdy}}$ of BCFT. Additionally, $G_N^{\left( {d + 1} \right)}$ is the gravitational constant in the d+1-dimensional holographic bulk spacetime, and L is the curvature radius of the bulk spacetime. We can also use $c = \frac{{3L}}{{2G_N^{\left( {d + 1} \right)}}}$ to express these entropies as functions of the central charge c of the d-dimensional field theory.
Furthermore, we have used
\be\label{srt}S(\tilde R) = \frac{1}{{4G_N^{\left( {d + 1} \right)}}}{d_{FO}} = \frac{{2L}}{{4G_N^{\left( {d + 1} \right)}}}\ln \frac{l}{\varepsilon },\ee
\be\label{sit}S(\tilde I) = \frac{1}{{4G_N^{\left( {d + 1} \right)}}}{d_{OE}} = \frac{L}{{4G_N^{\left( {d + 1} \right)}}}\ln \frac{l}{{\varepsilon \sin \theta }},\ee
and
\be S\left( R \right) = S\left( {R\tilde R} \right) = \frac{L}{{4G_N^{\left( {d + 1} \right)}}}\log \frac{\Lambda }{\varepsilon },\ee
\be S\left( I \right) = \frac{L}{{4G_N^{\left( {d + 1} \right)}}}\ln \frac{\Lambda }{{l\sin \theta }},\ee
where $\Lambda $ is the IR cutoff inside the bulk.

Let's make some comments on the coincidence (\ref{coi}). In the current context, all geometrically dual quantities in quantum information theory actually depend on the specific location of E (\ref{eco}). Remember that the point E is determined by the variation according to (\ref{bcft}), in other words, the point E is determined by the optimization program of ``finding’’ the island. On the other hand, it is not a priori for the half conditional mutual information ${F_{R\tilde I}}$ connecting the region R and $\tilde I$ to be equal to the half conditional mutual information ${F_{\tilde RI}}$ connecting the region $\tilde R$ and I, as different geometric surfaces are used in their computation. Using (\ref{bcft}) to find the position of the island automatically gives the flux ${F_{R\tilde I}}$ equal to ${F_{\tilde RI}}$!

\subsection{Connection Between the EWCS associated with the Island and Perfect Entanglement }
\begin{figure}
    \centering
    \includegraphics{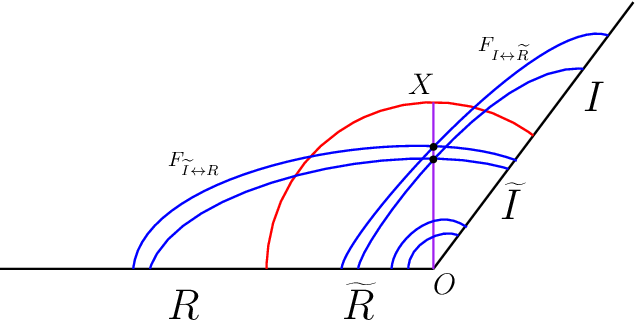}
    \caption{The threads in each bundle are present schematically. Note that for convenient we have omitted the other three thread bundles that do not contribute to the intrinsic correlation between $\tilde R$ and $\tilde I$.}
    \label{fig3.3.1}
\end{figure}

What does (\ref{coi}) imply? It brings to mind the relationship we reviewed in Section\ref{sec22} regarding the connection between two equal fluxes and the area of an EWCS. An inspiring conjecture is that the result $F_{R\tilde I} = F_{\tilde RI}$ is closely related to the EWCS (as shown in Fig.~\ref{fig3.3.1}) that characterizes the intrinsic correlation between $\tilde R$ and $\tilde I$. But let us firstly clarify the similarities and differences between these two scenarios. The similarity is evident: in Section\ref{sec22}, there exists an optimized point $H_*$ such that the number of threads connecting $A_1$ and $B_2$, denoted as ${F_{{A_1}{B_2}}}$, is equal to the number of threads connecting $A_2$ and $B_1$, and these two sets of threads are ``crossing’’. In the current scenario, there exists an optimized point $E$ that similarly makes the thread number $F_{\tilde RI}$ connecting the region $\tilde R$ and region $I$ exactly equal to the number of threads connecting $R$ and $\tilde I$, and these two sets of threads are also ``crossing’’ (see Fig.~\ref{fig3.3.1}). Thus, applying an analogy, if $F_{{A_1}{B_2}} = F_{{A_2}{B_1}}$ contributes precisely to the area of EWCS ${\sigma _{{A_1}:{A_2}}}$ characterizing the intrinsic correlation between $A_1$ and $A_2$ (as in (\ref{sig})), similarly, $F_{\tilde RI} = F_{R\tilde I}$ might contribute precisely to the area of EWCS $ \sigma _{\tilde R:\tilde I}$ characterizing the intrinsic correlation between $\tilde R$ and $\tilde I$! However, we must note a crucial difference. In Section\ref{sec22}, we artificially adjusted the point $H$ to make $F_{{A_1}{B_2}} = F_{{A_2}{B_1}}$ hold. In the current scenario, we follow the formula (\ref{bcft}), essentially searching for the position of the island, and coincidentally, this optimization procedure gives the result $F_{\tilde RI} = F_{R\tilde I}$.

Therefore, we arrive at a situation that is entirely different a priori, and a nontrivial new validation is needed regarding the relationship between fluxes and the area of EWCS. As shown in Fig.~\ref{fig3.3.1}, to define the EWCS $ {\sigma _{\tilde R:\tilde I}}$, we first define the semiclassical entanglement wedge $ W(\tilde R\tilde I) $, given by the region bounded by $\widetilde{I}$, $\widetilde{R}$, and the RT surface (the blue curve) of $\widetilde{I}\cup\widetilde{R}$. Then the EWCS $ {\sigma _{\tilde R:\tilde I}} $ is defined as seeking a minimal area extremal surface that partitions $ W(\tilde R\tilde I) $, with one half entirely touching $\tilde R$ and the other half entirely touching $\tilde I$. In other words, we need to find a minimal surface with one end at point $O$ and the other end on the RT surface—namely, the geodesic connecting points $E$ and $F$. The area of such an EWCS measures the intrinsic correlation (such as EoP) between $\tilde R$ and $\tilde I$. Let us parameterize the coordinates of endpoint of EWCS on the RT surface as $X = \left( { - l\cos \phi , l\sin \phi } \right)$, then the length of the geodesic between $O$ and $X$ must satisfy:
\be\frac{{l^2{{\cos }^2}\phi + {{(l\sin \phi - \varepsilon )}^2}}}{{2\varepsilon l\sin \phi }} + 1 = \cosh \frac{{d_{OX}}}{L}.\ee
Differentiating this equation with respect to $\phi $, we get the extreme value point:
\be d'_{OX} = 0 \Rightarrow \cos \phi = 0 \Rightarrow \phi = \frac{\pi }{2},\ee
and
\be 0 < \phi < \frac{\pi }{2}, \;\; d'_{OX} < 0,\ee
\be \frac{\pi }{2} < \phi < \pi - \theta, \;\; d'_{OX} > 0,\ee
Therefore, when $\phi = \frac{\pi }{2}$, $d_{OX}$ takes the minimum value. Finally, we can calculate the area of the EWCS as:
\be\label{area}\frac{{Area\left( {{\sigma _{\tilde R:\tilde I}}} \right)}}{{4G_N^{\left( {d + 1} \right)}}} = \frac{L}{{4G_N^{\left( {d + 1} \right)}}}\ln \frac{l}{\varepsilon}.\ee
In fact, it can be verified that, at this point, the geodesic connecting $O$ and $X$ is exactly a straight line in coordinate space, i.e., $x = 0$.

Now, to verify our conjecture, we should determine the number of threads connecting $\tilde R$ and $\tilde I$. Applying (\ref{sir}), (\ref{srt}), and (\ref{sit}), we can determine ${F_{\tilde R\tilde I}}$:
\be\label{mut}{F_{\tilde R\tilde I}} = {\textstyle{1 \over 2}}\left( {S\left( {\tilde I} \right) + S\left( {\tilde R} \right) - S\left( {\tilde I\tilde R} \right)} \right) = \frac{L}{{8G_N^{\left( {d + 1} \right)}}}\ln \frac{{l^2}}{{2{\varepsilon ^2}(1 + \cos \theta )}}.\ee
Thus, combining (\ref{rti}) and (\ref{mut}), we find that
\be{F_{R\tilde I}} + {F_{\tilde R\tilde I}} = \frac{L}{{4G_N^{\left( {d + 1} \right)}}}\ln \frac{l}{\varepsilon}.\ee
Therefore, from (\ref{area}), we find that 
\be\label{prop}\frac{{Area\left( {{\sigma _{\tilde R:\tilde I}}} \right)}}{{4G_N^{\left( {d + 1} \right)}}} = {F_{R\tilde I}} + {F_{\tilde R\tilde I}},\ee
thus confirming our conjecture!

Let us make some comments on (\ref{prop}). Firstly, from the viewpoint of mathematical analogy, the expression (\ref{prop}) can be viewed as a generalization of the BPE method~\cite{Wen:2021qgx,Camargo:2022mme,Wen:2022jxr} to holographic BCFT. On the other hand, from a physical viewpoint, it reveals a very noteworthy and insightful insight: in the semiclassical picture, an island $I$ and the part of ``radiation’’ $R$ should be characterized by perfect tensor entanglement, at least at a coarse-grained level. As pointed out in~\cite{yiyu2023}, only by endowing perfect tensor entanglement to two bundles of ``crossing’’ threads (such as ${F_{R\tilde I}}$ and ${F_{\tilde RI}}$), one of the thread bundle fluxes (such as ${F_{\tilde RI}}$) can be interpreted as contributing to the intrinsic correlation between two parts ($\tilde R$ and $\tilde I$ here) in a mixed state.

Let us clarify why ${F_{\tilde RI}} = {F_{R\tilde I}}$ together with (\ref{prop}) must imply the introduction of perfect entanglement. As shown in Fig.~\ref{fig3.3.1}, we depict a schematic diagram of thread configurations, wherein the threads in each bundle are present schematically. This thread configuration at a coarse-grained level characterizes the entanglement structure of the system $R \cup \tilde R \cup \tilde I \cup I$. If we simply assign an interpretation of Bell states (\ref{qut}) characterizing bipartite entanglement to each thread, and correspondingly, represent the thread configuration simply as a quantum state given by the direct product of all Bell states (\ref{coar}), then according to~\cite{yiyu2023}, by taking partial trace of this coarse-grained state, one can correctly obtain the entanglement entropies of all connected regions. However, to characterize the intrinsic correlation between $\tilde R$ and $\tilde I$, a more nontrivial structure is needed, even at the coarse-grained level. In the holographic duality, there are many quantum information theory quantities characterizing the correlation between two parts of a mixed state system. Here, we choose to use the entanglement of purification $ {E_P}(\tilde R:\tilde I) $~\cite{Nguyen:2017yqw,Takayanagi:2017knl} to characterize this correlation. The method is as follows: imagine introducing two auxiliary systems, denoted as $X$ and $Y$, such that $\tilde R \cup X \cup \tilde I \cup Y$ as a whole is in a pure state ${\left| \psi \right\rangle _{\tilde RX\tilde IY}}$. In this way, one can legitimately define the entanglement entropy between $\tilde RX$ and $\tilde IY$. Since there are infinitely many purification schemes, the entanglement of purification takes the minimum entanglement entropy between $\tilde RX$ and $\tilde IY$ as the correct intrinsic correlation between $\tilde R$ and $\tilde I$, namely,
\be\label{eop} {E_P}(\tilde R:\tilde I) = \mathop {\min }\limits_{{{\left| \psi \right\rangle }_{\tilde RX\tilde IY}}} S(\tilde RX).\ee
Similar to the RT formula, it has been proposed in~~\cite{Nguyen:2017yqw,Takayanagi:2017knl} that the EoP can be calculated by the area of the dual EWCS surface, i.e.,
\be {E_P}(\tilde R:\tilde I) = \frac{{{\rm{Area}}\left( {{\sigma _{\tilde R:\tilde I}}} \right)}}{{4G_N^{\left( {d + 1} \right)}}}.\ee
The key point is that if only bipartite entanglement is involved, the result, according to (\ref{coar}), of the EoP calculated by (\ref{eop}) is insufficient to give the exact value of the area of the EWCS. This can be understood intuitively as follows. As shown in Fig.~\ref{state}, remembering that the endpoints of the threads represent qutrits (\ref{qut}), when only bipartite entanglement is present, one can easily construct a purification scheme for mixed state system $\tilde R \tilde I$: identify the set of qudits on $I$ which are connected to $\tilde R$ by threads as the auxiliary system $X$, and identify the set of qutrits on $R$ which are connected to $\tilde I$ by threads as $Y$. This provides an optimal purification scheme that leads to
\be {E_P}(\tilde R:\tilde I) = {S_{opt}}(\tilde RX) = {F_{\tilde R\tilde I}}.\ee
Comparing with (\ref{prop}), we see that it is not enough to provide the correct value of the area of the EWCS. Essentially, this is because when using a coarse-grained state containing only bipartite entanglement, only the amount of entanglement exactly equal to the number of threads directly connecting $\tilde R$ and $\tilde I$ (${F_{\tilde R\tilde I}}$) characterizes the intrinsic correlation between $\tilde R$ and $\tilde I$.

On the other hand, as proposed in~\cite{yiyu2023}, as shown in Fig.~\ref{fig3.3.1}, if we couple each thread connecting $\tilde R$ and $I$ one-to-one with each thread connecting $R$ to $\tilde I$ to form a perfect tensor state (\ref{a1b2}), then the optimal purification scheme for measuring the intrinsic correlation between $\tilde R$ and $\tilde I$ will be as follows: identify the union of qudits on $I$ that are connected to $\tilde R$ via threads and qutrits on $R$ that are connected to $\tilde I$ via threads as Y, and identify X as an empty set. In this way, each perfect tensor ``knot’’ will contribute precisely a log3 to the total EoP, and since the threads in the two intersecting bundles are paired one-to-one, we will obtain ${F_{\tilde RI}}$ times the contribution. Finally, adding the contribution of bipartite entanglement directly connecting $\tilde R$ and $\tilde I$, i.e., ${F_{\tilde R\tilde I}}$, we obtain the complete (\ref{prop}) result.

\section{Computations for Various Scenarios}\label{sec4}


\begin{figure}[htbp]
\centerline{\includegraphics[scale=1]{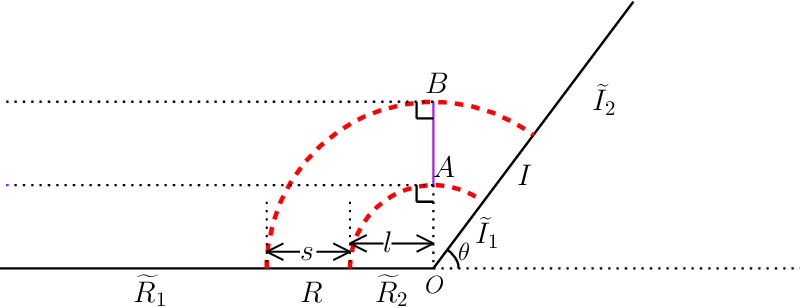}}
\caption{BCFT on an infinite straight line. The red dashed curves are RT surfaces corresponding to the semi-classical entropies in the semi-classical picture, and the purple curve is the EWCS for  $R\cup I$, which in this context characterizes the intrinsic correlation between a subregion R in CFT and the island I.}
\label{fig0}
\end{figure}
\subsection{BCFT on an Infinite Straight Line: finite region}\label{4.1}

In this subsection, we consider a little bit more complicate case shown in Figure.\ref{fig0}. 
As before, we express the number of threads in terms of the corresponding entanglement entropies, namely, via (\ref{cmi}),
\begin{equation}
\begin{split}
    F_{I\widetilde{R}_{1}}+F_{I\widetilde{R}_{2}}+F_{IR}&=\frac{1}{2}I\left(M;I|\widetilde{I_1}\right)\\
    &=\frac{1}{2}\left(S\left(M\cup\widetilde{I_1}\right)+ S\left(I\cup\widetilde{I_1}\right)- S\left(M\cup I\widetilde{I_1}\right)- S\left(\widetilde{I_1}\right)\right)\\
    &=\frac{1}{2}\left(S\left(I\widetilde{I_1}\right)+ S\left(I\widetilde{I_2}\right)- S\left(\widetilde{I_1}\right)- S\left(\widetilde{I_2}\right)\right),
\end{split}
\end{equation}
\begin{equation}
\begin{split}
    F_{\widetilde{I}_{1}R}+F_{\widetilde{I}_{2}R}+F_{IR}&=\frac{1}{2}I\left(Q;R|\widetilde{R_1}\right)\\
    &=\frac{1}{2}\left(S\left(Q\cup\widetilde{R_1}\right)+ S\left(R\cup\widetilde{R_1}\right)- S\left(Q\cup R\widetilde{R_1}\right)- S\left(\widetilde{R_1}\right)\right)\\
    &=\frac{1}{2}\left(  S\left(R\widetilde{R_1}\right)+ S\left(R\widetilde{R_2}\right)- S\left(\widetilde{R_1}\right)- S\left(\widetilde{R_2}\right)\right),
\end{split}
\end{equation}
where $M=\widetilde{R_1}R\widetilde{R_2}$ and  $Q=\widetilde{I_1}I\widetilde{I_2}$.
By using the RT formula, the semi-classical entropies above are found holographically to be

\begin{equation}
    S\left(\widetilde{I}_{1}\right)=\frac{L}{4G}\text{ln}\left(\frac{\Lambda}{\left(\ l+s\right)\text{sin}\theta}\right), \quad
S\left(\widetilde{I_{2}}\right)=\frac{L}{4G}\text{ln}\left(\frac{l}{\varepsilon\text{sin}\theta}\right),
\end{equation}

\begin{equation}
    S\left(\widetilde{R}_{1}\right)= S\left(R\widetilde{R}_{1}\right)=\frac{L}{4G}\text{ln}\left(\frac{\Lambda}{\varepsilon}\right), \quad
      S\left(\widetilde{R}_{2}\right)=\frac{L}{4G}\text{ln}\left(\frac{l}{\varepsilon}\right)^2
\end{equation}

\begin{equation}
     S\left(R\widetilde{R}_{2}\right)=\frac{L}{4G}\text{ln}\left(\frac{l+s}{\varepsilon}\right)^2,\quad S\left(I\widetilde{I}_{1}\right)=\frac{L}{4G}\text{ln}\left(\frac{\Lambda}{l\text{sin}\theta}\right)
\end{equation}

\begin{equation}
     S\left(I\widetilde{I}_{2}\right)=\frac{L}{4G}\text{ln}\left(\frac{l+s}{\varepsilon \text{sin}\theta}\right)
\end{equation}
From the results above, it is straightforward to see 
\begin{equation}\label{eq:9}
 F_{I\widetilde{R}_{1}}+F_{I\widetilde{R}_{2}}= F_{\widetilde{I}_{1}R}+F_{\widetilde{I}_{2}R},
\end{equation}
which is exactly the same situation as (\ref{coi}), that is, the number of threads (i.e., the half CMI) relating to the island $I$ and the CFT complement ${\tilde R}$ (which is $\tilde R \equiv {{\tilde R}_1} \cup {{\tilde R}_2}$ in this context) is coincidently equal to that relating to the subregion $R$ and the ``gravitational complement'' of the island (i.e., $\tilde I \equiv {{\tilde I}_1} \cup {{\tilde I}_2}$). Again, we expect that this implies the existence of perfect-tensor entanglement between $I$, $\widetilde{I}$, $R$, and $\widetilde{R}$.

To completely confirm our conjecture that the optimization procedure of searching for the position of the island automatically leads to the perfect entanglement pattern, we should study the EWCS (the purple line shown in Figure \ref{fig0}) for this case.
The length of the EWCS  $d_{AB}$  satisfies 
\begin{equation} \label{eq:5}
\begin{split}
\text{cosh}\left( \frac{d_{AB}}{L} \right)& =  1+\frac{(x_A-x_B)^2+(z_A-z_B)^2}{2z_Az_B} \\
 & =1+\frac{s^2}{2\left( l+s \right)s}\\
 & =\frac{1}{2}\left( \frac{l}{l+s}+\frac{l+s}{l} \right),
\end{split}
\end{equation}
so
\begin{equation}\label{eq:6}
    \frac{ d_{AB}}{4G}= \frac{L}{4G}\text{Log}\left( \frac{l+s}{l} \right). 
\end{equation}
On the other hand,
\begin{equation}\label{eq:1}
\begin{split}
   F_{\widetilde{I}_{1}R}+F_{\widetilde{I}_{2}R}+F_{IR}&=\frac{1}{2}\left(  S\left(R\widetilde{R_1}\right)+ S\left(R\widetilde{R_2}\right)- S\left(\widetilde{R_1}\right)- S\left(\widetilde{R_2}\right)\right)\\
   &=\frac{L}{4G}\text{Log}\left( \frac{l+s}{l} \right). 
\end{split}
\end{equation}
With eq.(\ref{eq:6}) and eq.(\ref{eq:1}), we indeed obtain the expecting relation between the number of threads and the length(area) of the corresponding EWCS, i.e.
\begin{equation}\label{eq:10}
    \frac{ d_{AB}}{4G}= F_{\widetilde{I}_{1}R}+F_{\widetilde{I}_{2}R}+F_{IR}.
\end{equation}
This again supports our interpretation of perfect entanglement pattern at the coarse-grained level.

For the readers familiar with the concept of partial entanglement entropy~\cite{Vidal:2014aal}, here we highlight anther comment: (\ref{eq:10}) also tells us: the area of EWCS $d_{AB}$ exactly provides the contribution of the degrees of freedom on $Q$ to the semi-classical entropy of region $R$. In other words, $d_{AB}$ exactly provides the semi-classical partial entangment entropy ${s_M}\left( R \right)$~\cite{Lin:2022aqf}, i.e., the contribution of $R$ to the semi-classical entropy of $M$.

\subsection{BCFT on a Finite disk: Connected Case}\label{sec42}
\begin{figure}[htbp]
\centerline{\includegraphics[scale=0.7]{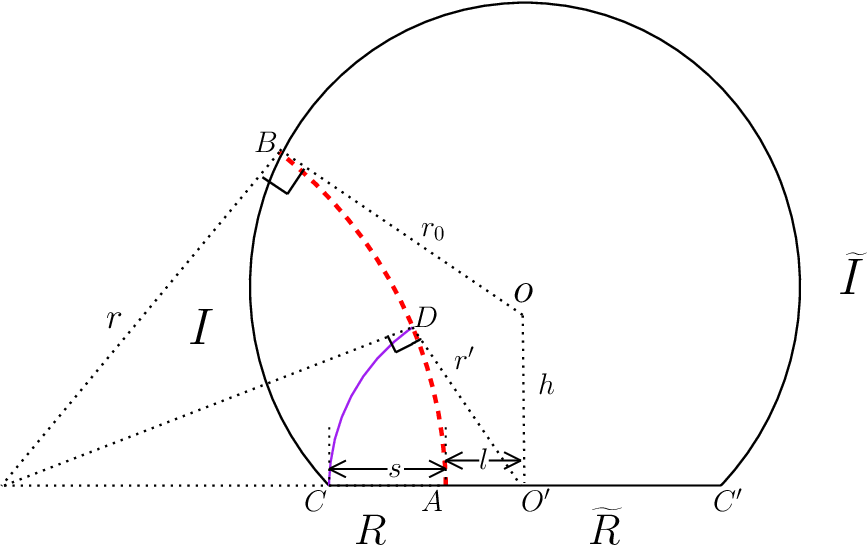}}
\caption{BCFT on a finite disk. The red dashed curves are RT surfaces corresponding to the semi-classical entropies in the semi-classical picture, and the purple curve is the EWCS for  $R\cup I$.}
\label{fig1}
\end{figure}

Another interesting example is the case where BCFT lives on a disk. For this, denote the coordinates except for the radial coordinate $z$ as ${X^\mu } = \left( {\tau ,x,\vec \chi } \right)$, where $\tau  = it$ is the Euclidean time. Then applying the following conformal map (where ${c^\mu }$ are arbitrary constants)~\cite{Fujita:2011fp,Berenstein:1998ij} 
\be
	{{X'}_\mu } = \frac{{{X_\mu } + {c_\mu }{X^2}}}{{1 + 2\left( {c \cdot X} \right) + {c^2} \cdot {X^2}}},\quad 
	z' = \frac{z}{{1 + 2\left( {c \cdot X} \right) + {c^2} \cdot {X^2}}}.
\ee
and performing a proper translation, we can map the BCFT on the half space defined by $x < 0$ to a BCFT living on a $d$ dimensional ball with radius $ k$, defined by
\be\label{rb}{\tau ^2} + {x^2} + {\vec \chi ^2} \le k^2.\ee
In this way, we can find the $Q$ brane as
\be\label{qb} {\tau ^2} + {x^2} + {\vec \chi ^2} + {\left( {z - {r_B}\sinh \frac{{{\rho _*}}}{L}} \right)^2} = k^2{\left( {\cosh \frac{{{\rho _*}}}{L}} \right)^2},\ee
which is also a sphere, as shown in figure~.

As before, to study the entanglement pattern, we calculate the number of threads between  $I$, $\widetilde{I}$, $R$, and $\widetilde{R}$ shown in Figure.\ref{fig1}, i.e.,
\begin{equation}\label{eq4.2.1}
\begin{split}
F_{I\widetilde{R}}+F_{IR}&=\frac{1}{2}I\left(I;M\right)\\
&=\frac{1}{2}\left(S\left(I\right)+S\left(M\right)-S\left(I\cup M\right)\right)\\
&=\frac{1}{2}\left(S\left(I\right)+S\left(I\widetilde{I}\right)-S\left(\widetilde{I}\right)\right),
\end{split}
\end{equation}
\begin{equation}\label{eq4.2.2}
\begin{split}
F_{\widetilde{I}R}+F_{IR}&=\frac{1}{2}I\left(R;Q\right)\\
&=\frac{1}{2}\left(S\left(R\right)+S\left(Q\right)-S\left(R\cup Q\right)\right)\\
&=\frac{1}{2}\left(S(R)+S\left(R\widetilde{R}\right)-S\left(\widetilde{R}\right)\right),
\end{split}
\end{equation}
where $M=R\widetilde{R}$ and $Q=I\widetilde{I}$.
The entanglement entropies in (\ref{eq4.2.1}) and (\ref{eq4.2.2}) are found as follow 
\begin{equation}\label{eq:17}
    S_{I}=\frac{L}{4G}\text{ln}\left(\frac{\left(x_{B}-x_{C}\right)^2+z_{B}^2}{z_{B}\varepsilon}\right),\quad
     S_{\widetilde{I}}=\frac{L}{4G}\text{ln}\left(\frac{\left(x_{B}+x_{C}\right)^2+z_{B}^2}{z_{B}\varepsilon}\right),
\end{equation}

\begin{equation}
    S_{R}=\frac{L}{4G}\text{ln}\left(\frac{\left(x_{A}-x_{C}\right)^2}{\varepsilon^2}\right),\quad
        S_{\widetilde{R}}=\frac{L}{4G}\text{ln}\left(\frac{\left(x_{A}+x_{C}\right)^2}{\varepsilon^2}\right)
\end{equation}
 \begin{equation}\label{eq:16}
     S_{I\widetilde{I}}= S_{R\widetilde{R}}=\frac{L}{4G}\text{ln}\left(\frac{\left(2x_C\right)^2}{\varepsilon^2}\right).
 \end{equation}
With basic geometry, we find that the coordinates above are such that
\begin{equation}\label{eq:13}
   x_B=\frac{2l\left(h-r_{0}\right)r_{0}}{l^2+\left(h-r_{0}\right)^2},\quad 
   z_B=h+r_{0}-\frac{2l^2r_{0}}{l^2+(h-r_{0})^2},
\end{equation}
where $r_0=\sqrt{h^2+\left(l+s\right)^2}$
, and
\begin{equation}\label{eq:12}
 x_A=-l ,\quad 
x_C=-\left(l+s\right).
\end{equation}
Therefore
\begin{equation}
   F_{I\widetilde{R}}+F_{IR}=\frac{L}{4G}\text{ln}\left(\frac{2s(l+s)}{(2l+s)\varepsilon}\right),
\end{equation}
\begin{equation}\label{eq4.2.3}
   F_{\widetilde{I}R}+F_{IR}=\frac{L}{4G}\text{ln}\left(\frac{2s(l+s)}{(2l+s)\varepsilon}\right).
\end{equation}
From the two equations above, we then obtain
\begin{equation}
    F_{I\widetilde{R}}= F_{\widetilde{I}R},
\end{equation}
again, the positioning of the entanglement island automatically leads to a perfect-tensor entanglement pattern.

Next, we study the EWCS $CD$ shown in Figure \ref{fig1}.
The coordinates for $D$ are 
\begin{equation}
    z_{D}=\frac{rr'}{r+l},\quad x_D=\sqrt{r'^2-z_D^2},
\end{equation}
where $r'=\sqrt{(r+l)^2-r^2}$ and $r=(-r_{0}^2+l^2+h^2)/(2l)$.
Then the length of the EWCS $CD$ is found to be
\begin{equation}\label{eq4.2.4}
    \frac{d_{CD}}{4G}=\frac{L}{4G}\text{ln}\left(\frac{\left(x_{D}-x_{C}\right)^2+z_{D}^2}{z_{D}\varepsilon}\right)=\frac{L}{4G}\text{ln}\left(\frac{2s(l+s)}{(2l+s)\varepsilon}\right).
\end{equation}
From eq.(\ref{eq4.2.3}) and eq.(\ref{eq4.2.4}), we get
\begin{equation}
    \frac{d_{CD}}{4G} =F_{\widetilde{I}R}+F_{IR},
\end{equation}
as expected.

\subsection{BCFT on a Finite Disk: Disconnected Case}\label{sec43}
\begin{figure}[htbp]
\centerline{\includegraphics[scale=0.7]{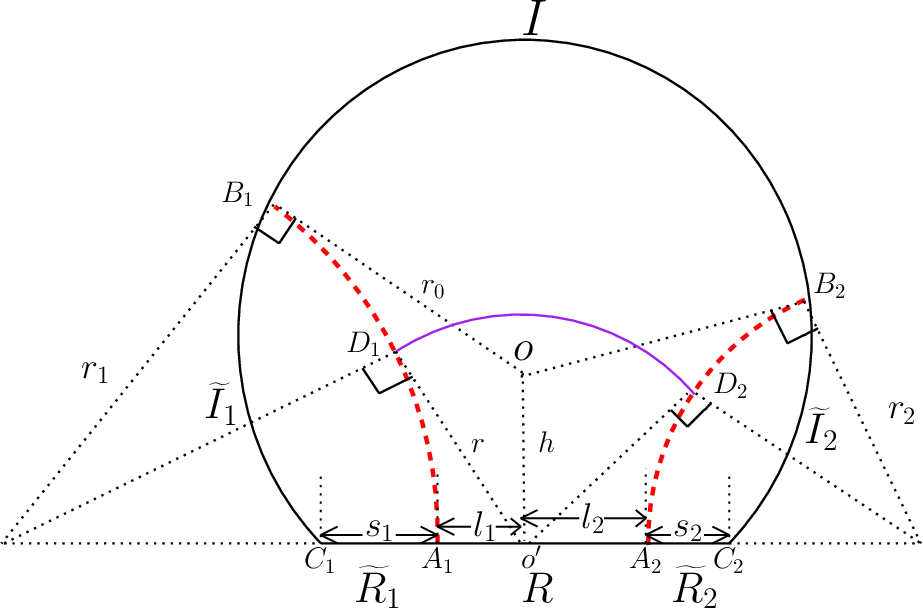}}
\caption{BCFT on a finite disk. The red dashed curves are RT surfaces corresponding to the semi-classical entropies in the semi-classical picture, and the purple curve is the EWCS for  $R\cup I$.}
\label{fig2}
\end{figure}
In this subsection, we consider a region $R$ shown in figure \ref{fig2}. 
Similar to what we have in subsection \ref{4.1}, the corresponding numbers of threads are
\begin{equation}
\begin{split}
    F_{I\widetilde{R}_{1}}+F_{I\widetilde{R}_{2}}+F_{IR}&=\frac{1}{2}I\left(M;I|\widetilde{I_1}\right)\\
    &=\frac{1}{2}\left(S\left(M\cup\widetilde{I_1}\right)+ S\left(I\cup\widetilde{I_1}\right)- S\left(M\cup I\widetilde{I_1}\right)- S\left(\widetilde{I_1}\right)\right)\\
    &=\frac{1}{2}\left(S\left(I\widetilde{I_1}\right)+ S\left(I\widetilde{I_2}\right)- S\left(\widetilde{I_1}\right)- S\left(\widetilde{I_2}\right)\right),
\end{split}
\end{equation}
\begin{equation}
\begin{split}
    F_{\widetilde{I}_{1}R}+F_{\widetilde{I}_{2}R}+F_{IR}&=\frac{1}{2}I\left(Q;R|\widetilde{R_1}\right)\\
    &=\frac{1}{2}\left(S\left(Q\cup\widetilde{R_1}\right)+ S\left(R\cup\widetilde{R_1}\right)- S\left(Q\cup R\widetilde{R_1}\right)- S\left(\widetilde{R_1}\right)\right)\\
    &=\frac{1}{2}\left(  S\left(R\widetilde{R_1}\right)+ S\left(R\widetilde{R_2}\right)- S\left(\widetilde{R_1}\right)- S\left(\widetilde{R_2}\right)\right),
\end{split}
\end{equation}
where $M=\widetilde{R_1}R\widetilde{R_2}$ and  $Q=\widetilde{I_1}I\widetilde{I_2}$.
By using RT formula, the entanglement entropies above are computed to be
\begin{equation}\label{eq7}
    S\left(\widetilde{R}_{1}\right)=\frac{L}{4G}\text{ln}\left(\frac{s_1}{\varepsilon}\right)^2,\quad
    S\left(\widetilde{R}_{2}\right)=\frac{L}{4G}\text{ln}\left(\frac{s_2}{\varepsilon}\right)^2,
\end{equation}

\begin{equation}
    S\left(\widetilde{I}_{1}\right)=\frac{L}{4G}\text{ln}\left(\frac{\left(x_{B_{1}}+s_1+l_1\right)^2+z_{B_{1}}^2}{\varepsilon z_{B_{1}}}\right),\quad
     S\left(\widetilde{I}_{2}\right)=\frac{L}{4G}\text{ln}\left(\frac{\left(x_{B_{2}}-s_2-l_2\right)^2+z_{B_{2}}^2}{\varepsilon z_{B_{2}}}\right),
\end{equation}

\begin{equation}\label{eq8}
    S\left(R\widetilde{R}_{1}\right)=\frac{L}{4G}\text{ln}\left(\frac{s_1+l_1+l_2}{\varepsilon}\right)^2,\quad
      S\left(R\widetilde{R}_{2}\right)=\frac{L}{4G}\text{ln}\left(\frac{s_2+l_1+l_2}{\varepsilon}\right)^2,
\end{equation}
\begin{equation}
    S\left(I\widetilde{I}_{1}\right)=\frac{L}{4G}\text{ln}\left(\frac{\left(x_{B_{2}}+s_1+l_1\right)^2+z_{B_{2}}^2}{\varepsilon z_{B_{2}}}\right),\quad
     S\left(I\widetilde{I}_{2}\right)=\frac{L}{4G}\text{ln}\left(\frac{\left(x_{B_{1}}-s_2-l_2\right)^2+z_{B_{1}}^2}{\varepsilon z_{B_{1}}}\right).
\end{equation}
The coordinates for $B_1$ and $B_2$ are found as follow
\begin{equation}\label{eq5}
   x_{B_1}=\frac{2l_1\left(h-r_{0}\right)r_{0}}{l_1^2+\left(h-r_{0}\right)^2},\quad 
   z_{B_1}=h+r_{0}-\frac{2l_1^2r_{0}}{l_1^2+(h-r_{0})^2},
\end{equation}

\begin{equation}\label{eq6}
   x_{B_2}=-\frac{2l_2\left(h-r_{0}\right)r_{0}}{l_2^2+\left(h-r_{0}\right)^2},\quad 
   z_{B_2}=h+r_{0}-\frac{2l_2^2r_{0}}{l_2^2+(h-r_{0})^2}.
\end{equation}
With the results above and $r_0=\sqrt{(l_1+s_1)^2+h^2}=\sqrt{(l_2+s_2)^2+h^2}$, we find 
\begin{equation}
\begin{split}
     F_{I\widetilde{R}_{1}}+F_{I\widetilde{R}_{2}}+F_{IR}&=\frac{1}{2}\left(  S\left(I\widetilde{I_1}\right)+ S\left(I\widetilde{I_2}\right)- S\left(\widetilde{I_1}\right)- S\left(\widetilde{I_2}\right)\right)\\
     &=\frac{L}{4G}\text{ln}\left(\frac{(l_1+l_2+s_1)(l_1+l_2+s_2)}{s_1s_2}\right),
\end{split} 
\end{equation}

\begin{equation}\label{eq4.3.1}
\begin{split}
    F_{\widetilde{I}_{1}R}+F_{\widetilde{I}_{2}R}+F_{IR}&=\left(  S\left(R\widetilde{R_1}\right)+ S\left(R\widetilde{R_2}\right)- S\left(\widetilde{R_1}\right)- S\left(\widetilde{R_2}\right)\right)\\
    &=\frac{L}{4G}\text{ln}\left(\frac{(l_1+l_2+s_1)(l_1+l_2+s_2)}{s_1s_2}\right)
\end{split}
\end{equation}
Thus, we have
\begin{equation}
 F_{I\widetilde{R}_{1}}+F_{I\widetilde{R}_{2}}= F_{\widetilde{I}_{1}R}+F_{\widetilde{I}_{2}R}.
\end{equation}
Therefore in this case, again, the positioning of the entanglement island automatically leads to a perfect-tensor entanglement pattern.

The coordinates for $D_1$ and $D_2$ are 
\begin{equation}
    x_{D_1}=-\sqrt{r^2-z_{D_1}^2},\quad
    z_{D_1}=\frac{r_1r}{r_1+l_1}
\end{equation}

\begin{equation}
    x_{D_2}=\sqrt{r^2-z_{D_2}^2},\quad
    z_{D_2}=\frac{r_2r}{r_2+l_2},
\end{equation}
with
\begin{equation}
    r_1=\frac{r_0^2-l_1^2-h^2}{2l_1},\quad
    r_2=\frac{r_0^2-l_2^2-h^2}{2l_2},
\end{equation}
where $r=\sqrt{(r_1+l_1)^2-r_1^2}=\sqrt{(r_2+l_2)^2-r_2^2}$. Then the length of the EWCS $D_1D_2$ (the purple curve shown in figure \ref{fig2}) satisfies
\begin{equation}
\begin{split}
\text{cosh}\left( \frac{d_{D_1D_2}}{L} \right)& =  1+\frac{(x_{D_1}-x_{D_2})^2+(z_{D_1}-z_{D_2})^2}{2z_{D_1}z_{D_2}} \\
 & =1+\frac{2\left(l_1+l_2\right)^2\left(l_1+s_1\right)^2}{s_1\left(2l_1+s_1\right)\left(l_1-l_2+s_1\right)\left(l_1+l_2+s_1\right)}\\
 &=\frac{1}{2}\left(\frac{s_1(l_1-l_2+s_1)}{(2l_1+s_1)(l_1+l_2+s_1)}+\frac{(2l_1+s_1)(l_1+l_2+s_1)}{s_1(l_1-l_2+s_1)}\right),
\end{split}
\end{equation}
i.e.,
\begin{equation}\label{eq4.3.2}
\begin{split}
    \frac{d_{D_1D_2}}{4G}&=\frac{L}{4G}\text{ln}\left(\frac{(2l_1+s_1)(l_1+l_2+s_1)}{s_1(l_1-l_2+s_1)}\right)\\
    &=\frac{L}{4G}\text{ln}\left(\frac{(l_1+l_2+s_1)(l_1+l_2+s_2)}{s_1s_2}\right),
\end{split}
\end{equation}
where in the second line, we used $s_1+l_1=s_2+l_2$.
With eq.(\ref{eq4.3.1}) and eq.(\ref{eq4.3.2}), we find the expecting relation, i.e.,
\begin{equation}
    \frac{ d_{D_1D_2}}{4G}=F_{\widetilde{I}_{1}R}+F_{\widetilde{I}_{2}R}+F_{IR}.
\end{equation}
\subsection{Simulating the Black Hole Information Problem}\label{sec44}

\begin{figure}[htbp]
\centerline{\includegraphics[scale=0.7]{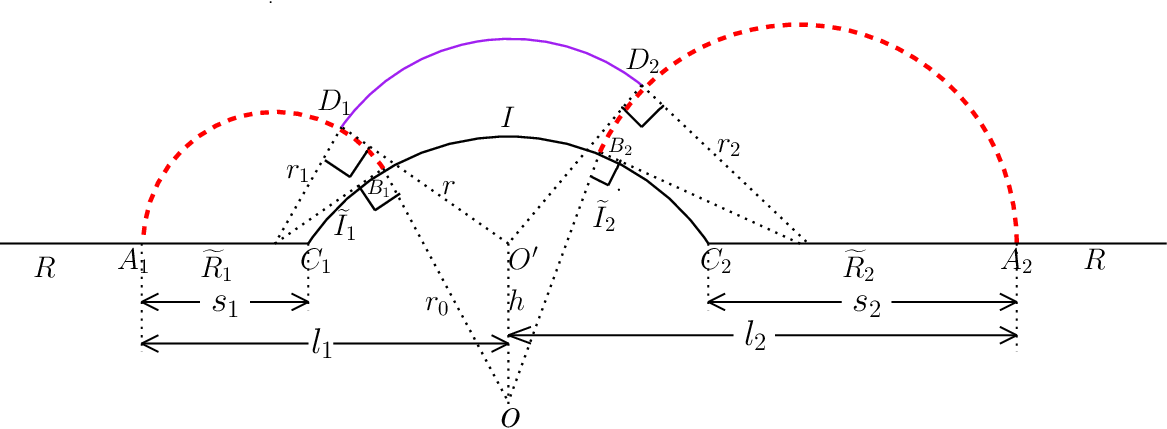}}
\caption{The holographic BCFT used for the simulation of Black Hole and Hawking radiation. The Brane (the curve between $C_1$ and $C_2$) and region $R$ play the roles of Black Hole and the corresponding Hawking radiation, respectively.}
\label{fig3}
\end{figure}

In this subsection we discuss one of the most direct and interesting examples in~\cite{Rozali:2019day} (see also~\cite{Sully:2020pza}), which simulated a two-sided black hole coupled to an auxiliary radiation system by a holographic BCFT system in the thermofield double state by applying the AdS/BCFT correspondence. This setup is similar to the one in figure~\ref{fig1} in which the boundary of a half plane is mapped to the boundary of a disk, except that now we map the half plane on which the BCFT lives to the Euclidean plane with the disk removed, as shown in figure~\ref{fig3}. It was argued that in the limit that the number of local degrees of freedom on the boundary of this BCFT is large compared to the number of local degrees of freedom in this bulk CFT itself, the ETW brane extending from the boundary of the disk can simulate a black hole because the brane itself has causal horizons. In this way this setup models a two-sided 2d black hole coupled to a pair of symmetrical auxiliary radiation systems. Note that this system is not an evaporating black hole, but one with the auxiliary radiation system has the same temperature as the black hole such that the two systems are in equilibrium. Furthermore, in a particular conformal frame, this system has a static energy density.~\footnote{Although there is no net energy exchange between the black hole and the radiation system in this setup, \cite{Rozali:2019day} argued that the information from black hole will ``escape'' (or be ``encoded'') into the radiation system $R$.} However, the calculation in~\cite{Rozali:2019day} reveals that, for a subsystem $R = \left( { - \infty ,\; - {x_0}} \right] \cup \left[ {{x_0},\;\infty } \right)$ consisting of the union of  two symmetric half-lines in each CFT, the entanglement entropy still evolves with time and undergoes a typical phase transition characterized by Page curve, similar to the ones discussed in~\cite{Almheiri:2019hni,Almheiri:2019psf,Penington:2019npb}. This phase transition is essentially because in AdS/BCFT correspondence, the RT surface calculating the true entanglement entropy of $R$ can be in an island phase, i.e., the RT surface can anchor on the ETW brane, as shown in figure~.

Now let us investigate this interesting and instructive setup in details from the viewpoint of coarse-grained state. As before, we calculate the ``flow fluxes'' characterizing the CMIs between different parts of this system:
\begin{equation}
\begin{split}
    F_{I\widetilde{R}_{1}}+F_{I\widetilde{R}_{2}}+F_{IR}&=\frac{1}{2}I\left(M;I|\widetilde{I_1}\right)\\
    &=\frac{1}{2}\left(S\left(M\cup\widetilde{I_1}\right)+ S\left(I\cup\widetilde{I_1}\right)- S\left(M\cup I\widetilde{I_1}\right)- S\left(\widetilde{I_1}\right)\right)\\
    &=\frac{1}{2}\left(S\left(I\widetilde{I_1}\right)+ S\left(I\widetilde{I_2}\right)- S\left(\widetilde{I_1}\right)- S\left(\widetilde{I_2}\right)\right),
\end{split}
\end{equation}
\begin{equation}
\begin{split}
    F_{\widetilde{I}_{1}R}+F_{\widetilde{I}_{2}R}+F_{IR}&=\frac{1}{2}I\left(Q;R|\widetilde{R_1}\right)\\
    &=\frac{1}{2}\left(S\left(Q\cup\widetilde{R_1}\right)+ S\left(R\cup\widetilde{R_1}\right)- S\left(Q\cup R\widetilde{R_1}\right)- S\left(\widetilde{R_1}\right)\right)\\
    &=\frac{1}{2}\left(  S\left(R\widetilde{R_1}\right)+ S\left(R\widetilde{R_2}\right)- S\left(\widetilde{R_1}\right)- S\left(\widetilde{R_2}\right)\right),
\end{split}
\end{equation}
where $M=\widetilde{R_1}R\widetilde{R_2}$ and  $Q=\widetilde{I_1}I\widetilde{I_2}$.
Via RT formula, we work out

\begin{equation}
    S\left(\widetilde{R}_{1}\right)=\frac{L}{4G}\text{ln}\left(\frac{s_1}{\varepsilon}\right)^2,\quad
    S\left(\widetilde{R}_{2}\right)=\frac{L}{4G}\text{ln}\left(\frac{s_2}{\varepsilon}\right)^2.
\end{equation}

\begin{equation}
    S\left(\widetilde{I}_{1}\right)=\frac{L}{4G}\text{ln}\left(\frac{\left(x_{B_{1}}+l_1-s_1\right)^2+z_{B_{1}}^2}{\varepsilon z_{B_{1}}}\right),\quad
     S\left(\widetilde{I}_{2}\right)=\frac{L}{4G}\text{ln}\left(\frac{\left(x_{B_{2}}-l_2+s_2\right)^2+z_{B_{2}}^2}{\varepsilon z_{B_{2}}}\right),
\end{equation}

\begin{equation}
    S\left(R\widetilde{R}_{1}\right)=\frac{L}{4G}\text{ln}\left(\frac{l_1+l_2-s_1}{\varepsilon}\right)^2,\quad
      S\left(R\widetilde{R}_{2}\right)=\frac{L}{4G}\text{ln}\left(\frac{l_1+l_2-s_2}{\varepsilon}\right)^2,
\end{equation}

\begin{equation}
    S\left(I\widetilde{I}_{1}\right)=\frac{L}{4G}\text{ln}\left(\frac{\left(x_{B_{2}}+l_1-s_1\right)^2+z_{B_{2}}^2}{\varepsilon z_{B_{2}}}\right),\quad
     S\left(I\widetilde{I}_{2}\right)=\frac{L}{4G}\text{ln} \left(\frac{\left(x_{B_{1}}-l_2+s_2\right)^2+z_{B_{1}}^2}{\varepsilon z_{B_{1}}}\right).
\end{equation}
By using basic geometry, we then work out the coordinates for $B_1$ and $B_2$:
\begin{equation}\label{eq454}
   x_{B_1}=\frac{2l_1\left(h-r_{0}\right)r_{0}}{l_1^2+\left(h-r_{0}\right)^2},\quad 
   z_{B_1}=h+r_{0}-\frac{2l_1^2r_{0}}{l_1^2+(h-r_{0})^2},
\end{equation}
\begin{equation}\label{eq456}
   x_{B_2}=-\frac{2l_2\left(h-r_{0}\right)r_{0}}{l_2^2+\left(h-r_{0}\right)^2},\quad 
   z_{B_2}=h+r_{0}-\frac{2l_2^2r_{0}}{l_2^2+(h-r_{0})^2},
\end{equation}
where $r_0=\sqrt{(l_1-s_1)^2+h^2}=\sqrt{(l_2-s_2)^2+h^2}$.
Using the results for the entanglement entropies and coordinates above, we find
\begin{equation}\label{eq4.4.1}
\begin{split}
     F_{I\widetilde{R}_{1}}+F_{I\widetilde{R}_{2}}+F_{IR}&=\frac{1}{2}\left(  S\left(I\widetilde{I_1}\right)+ S\left(I\widetilde{I_2}\right)- S\left(\widetilde{I_1}\right)- S\left(\widetilde{I_2}\right)\right)\\
     &=\frac{L}{4G}\text{ln}\left(\frac{(l_1+l_2-s_1)(l_1+l_2-s_2)}{s_1s_2}\right),
\end{split} 
\end{equation}
\begin{equation}\label{eq4.4.2}
\begin{split}
    F_{\widetilde{I}_{1}R}+F_{\widetilde{I}_{2}R}+F_{IR}&=\left(  S\left(R\widetilde{R_1}\right)+ S\left(R\widetilde{R_2}\right)- S\left(\widetilde{R_1}\right)- S\left(\widetilde{R_2}\right)\right)\\
    &=\frac{L}{4G}\text{ln}\left(\frac{(l_1+l_2-s_1)(l_1+l_2-s_2)}{s_1s_2}\right).
\end{split}
\end{equation}
Again! Eq.(\ref{eq4.4.1}) and eq.(\ref{eq4.4.2}) lead to a perfect entanglement pattern:
\begin{equation}
 F_{I\widetilde{R}_{1}}+F_{I\widetilde{R}_{2}}= F_{\widetilde{I}_{1}R}+F_{\widetilde{I}_{2}R}.
\end{equation}

Let us keep going forward to calculate the length of the EWCS 
 ${D_1D_2}$(the purple curve shown in Figure \ref{fig3}) in this context. 
The coordinates for $D_1$ and $D_2$ are found to be
\begin{equation}
    x_{D_1}=-\sqrt{r^2-z_{D_1}^2},\quad
    z_{D_1}=\frac{r_1r}{l_1-r_1}
\end{equation}

\begin{equation}
    x_{D_2}=\sqrt{r^2-z_{D_2}^2},\quad
    z_{D_2}=\frac{r_2r}{l_2-r_2},
\end{equation}
with
\begin{equation}
    r_1=\frac{l_1^2+h^2-r_0^2}{2l_1},\quad
    r_2=\frac{l_2^2+h^2-r_0^2}{2l_2},
\end{equation}
where $r=\sqrt{(l_1-r_1)^2-r_1^2}=\sqrt{(l_2-r_2)^2-r_2^2}$. Then we have
\begin{equation}
\begin{split}
\text{cosh}\left( \frac{d_{D_1D_2}}{L} \right)& =  1+\frac{(x_{D_1}-x_{D_2})^2+(z_{D_1}-z_{D_2})^2}{2z_{D_1}z_{D_2}} \\
 & =1-\frac{2\left(l_1+l_2\right)^2\left(l_1-s_1\right)^2}{s_1\left(2l_1-s_1\right)\left(l_1-l_2-s_1\right)\left(l_1+l_2-s_1\right)}\\
 &=\frac{1}{2}\left(\frac{s_1(l_2-l_1+s_1)}{(2l_1-s_1)(l_1+l_2-s_1)}+\frac{(2l_1-s_1)(l_1+l_2-s_1)}{s_1(l_2-l_1+s_1)}\right),
\end{split}
\end{equation}
i.e.,
\begin{equation}
\begin{split}\label{eq4.4.3}
  \frac{d_{D_1D_2}}{4G}&=\frac{L}{4G}\text{ln}\left(\frac{(2l_1-s_1)(l_1+l_2-s_1)}{s_1(l_2-l_1+s_1)}\right)\\
  &=\frac{L}{4G}\text{ln}\left(\frac{(l_1+l_2-s_1)(l_1+l_2-s_2)}{s_1s_2}\right),
\end{split}
\end{equation}
where in the second line we used $l_1-s_1=l_2-s_2$.
Somewhat surprisingly, eq.(\ref{eq4.4.2}) and eq.(\ref{eq4.4.3}) again give us
\begin{equation}
    \frac{ d_{D_1D_2}}{4G}= F_{\widetilde{I}_{1}R}+F_{\widetilde{I}_{2}R}+F_{IR}.
\end{equation}

Starting with a rather simple surmise, a series of precise and rather complex calculations have been carefully carried out. However, somewhat unexpected, the coincidence of the equal ``crossing fluxes" and the exact match of the EWCS with the sum of fluxes have not yet been broken. Although the picture of the Island/BCFT duality is not yet fully understood, this mathematical nontriviality will undoubtedly enhance the plausibility of investigating the BCFT and the brane-object in the semi-classical picture.


\section{Conclusion and Discussion}

This work was inspired by the recently established island/BCFT duality~\cite{Suzuki:2022xwv,Izumi:2022opi}, which suggests that the AdS/BCFT duality~\cite{Takayanagi:2011zk,Fujita:2011fp} can serve as a minimalist model for analyzing the surprising island effects that arise when considering the entanglement entropy of a subregion R in a field theory coupled to a gravity system in the semiclassical picture~\cite{Almheiri:2019hni,Penington:2019npb,Almheiri:2019psf}. We applied the idea of coarse-grained states~\cite{yiyu2023}, which are directly constructed from sets of CMIs, to study the entanglement structure between different parts in such setups at the coarse-grained level. Based on the understanding of the connection between the entanglement wedge cross-section and perfect tensor entanglement in~\cite{yiyu2023}, we discovered a very interesting phenomenon: in the semiclassical picture, the positioning of an entanglement island automatically gives rise to a pattern of perfect tensor entanglement between the subregion R and the island, and the sum of this perfect tensor entanglement contribution and the bipartite entanglement contribution precisely gives the area of the EWCS that characterizes the intrinsic correlation between R and I (or $\tilde R$ and $\tilde I$). The most noteworthy aspect of this phenomenon is that initially, we determined the boundary point between the island and its complement on the gravity brane according to the island rule (\ref{bcft}), and this does not a priori lead to the balanced result that ${F_{R\tilde I}} = {F_{\tilde RI}}$, i.e., the half CMI relating R and $\tilde I$ equals that relating $\tilde R$ and I. In a sense, this symmetry implies that the island I mirrors the information of the R region, while $\tilde I$ mirrors the information of the $\tilde R$ region. Although we examined only the simple case of island effects simulated by holographic BCFT (in 2d), this insight may also be applicable to more general island effects. This will be a research direction worth delving into in the future.

For the readers familiar with the concept of partial entanglement entropy~\cite{Vidal:2014aal}, our results also tells that the area of the EWCS relating to $R$ and the island $I$ exactly provides the contribution of the degrees of freedom on $Q$ to the semi-classical entropy of region $R$. In other words, this area exactly provides the semi-classical partial entangment entropy ${s_M}\left( R \right)$~\cite{Lin:2022aqf}, i.e., the contribution of $R$ to the semi-classical entropy of $M$.

Furthermore, it is worth noting that typically, R and I are envisioned to be connected through a wormhole in a higher dimension~\cite{Penington:2019kki,Almheiri:2019qdq}. In the context studied in this paper, the EWCS plays the role of the minimal area surface separating the ``channel’’ connecting R and I. The connection between these two concepts is also noteworthy~\cite{Bao:2018fso,Bao:2018zab,Bhattacharya:2020ymw}. In particular, we demonstrated that perfect tensor entanglement plays a crucial role in characterizing the duals of such minimal area surfaces. We believe this will contribute to a deeper understanding of the essence of the island. Finally, it can not be ignored that purely from a mathematical viewpoint, the core equation of this paper (\ref{prop}) can be viewed as a generalization of the so-called BPE method~\cite{Wen:2021qgx,Camargo:2022mme,Wen:2022jxr} in holographic BCFT. This, in a sense, demonstrates the universality of the BPE method in holographic duality, and contemplating the nature of this universality is also a topic worthy of further investigation.

\section*{Acknowledgement}
We would like to thank Ling-Yan Hung and Yuan Sun for useful discussions.

\end{document}